\documentclass[%
 reprint,
 superscriptaddress,
 amsmath,amssymb,
 aps,
 prl
 floatfix,
]{revtex4-1}

\usepackage{graphicx}
\usepackage{dcolumn}
\usepackage{bm}
\usepackage[dvipsnames]{xcolor}

\newcommand{\NBI}{Nanoscience Center, Niels Bohr Institute, University of Copenhagen, DK-2100 Copenhagen, Denmark}
\newcommand{\ILL}{Institut Laue-Langevin, 71 Avenue des Martyrs, 38000 Grenoble, France}
\newcommand{\risoe}{Department of Energy Conversion and Storage, Technical University of Denmark, Frederiksborgvej 399, DK-4000 Roskilde, Denmark}
\newcommand{\uconn}{ Department of Physics, University of Connecticut, Storrs, Connecticut 06269, USA}

\newcommand{\LSCO}{La$_{2-x}$Sr$_{x}$CuO$_4$}
\newcommand{\LSCOOsix}{La$_{1.94}$Sr$_{0.06}$CuO$_{4.035}$}
\newcommand{\LSCOO}{La$_{2-x}$Sr$_{x}$CuO$_{4+\delta}$}
\newcommand{\LCOO}{La$_2$CuO$_{4+\delta}$}
\newcommand{\LNSCO}{La$_{1.48}$Nd$_{0.4}$Sr$_{0.12}$CuO$_4$}
\newcommand{\LSCOTwenty}{La$_{1.80}$Sr$_{0.20}$CuO$_4$}
\newcommand{\LSCOopt}{La$_{1.85}$Sr$_{0.15}$CuO$_4$}
\newcommand{\LSCOseven}{La$_{1.93}$Sr$_{0.07}$CuO$_4$}
\newcommand{\LSCOTwentyfive}{La$_{1.75}$Sr$_{0.25}$CuO$_4$}

\newcommand{\LBCOtwelve}{La$_{1.875}$Ba$_{0.125}$CuO$_4$}

\newcommand{\Tc}{$T_\text{c}$}

\usepackage[normalem]{ulem}

\usepackage[colorinlistoftodos]{todonotes}

\begin{document}

\preprint{APS/123-QED}

\title{Anomalous Dispersion of LO Phonons in Oxygen-Doped La$_{2-x}$Sr$_{x}$CuO$_{4+\delta}$}

\author{Tim Tejsner}
\affiliation{\ILL}
\affiliation{\NBI}
\author{Andrea Piovano}
\affiliation{\ILL}
\author{Ana \c{T}u\c{t}ueanu}
\affiliation{\ILL}
\affiliation{\NBI}
\author{Astrid T. R\o{}mer}
\affiliation{\ILL}
\author{Barrett O. Wells}
\affiliation{\uconn}
\author{Jean-Claude Grivel}
\affiliation{\risoe}
\author{Martin Boehm}
\affiliation{\ILL}
\author{Linda Udby}
\affiliation{\NBI}


%

\date{\today}

\begin{abstract}

Inelastic neutron scattering has been used to study the in-plane Cu-O bond-stretching mode in oxygen doped \LSCOO{} ($T_c = 38\,\text{K}$) and \LCOO{} ($T_c = 43\,\text{K}$). Similar to results from optimally doped \LSCOopt{} ($T_c = 35\,\text{K}$), we observe anomalous features in the dispersion of this half-breathing mode in the form of a softening halfway through the Brillouin Zone. Considering the differences in electronic structure and local environment between the oxygen- and strontium-doped compounds with similar \Tc{}, we rule out a connection between the phonon anomaly and structural instabilities related to the specific dopant type. We interpret the phonon anomaly as a signature of correlated charge fluctuations possibly connected to stripes.
\end{abstract}

\maketitle


A widely accepted description of the low-temperature electronic state of underdoped cuprate superconductors is the `stripe'-picture, where, due to hole-doping, the anti-ferromagnetic (AFM) ground state of the parent compound is segregated by channels of charge resulting in magnetic anti-phase boundaries \cite{Tranquada1995}. Experimentally, the magnetic part of the stripes shows up in neutron scattering experiments as a modulated AFM structure (static magnetic stripes) and excitations emerging with similar wavevectors, in the low energy regime (dynamic magnetic stripes).

While experimental evidence of magnetic stripes have been extensively documented (see e.g. \cite{Tranquada2013} for a review), the charge component of the stripes is more elusive. Static charge stripes only show up in superconducting (SC) samples close to the $x=\frac{1}{8}$ anomaly \cite{Tranquada1995, Tranquada1996, Christensen2014, Thampy2014, Croft2014} and direct evidence of dynamic charge stripes has only been reported for isostructural, but insulating La$_{2-x}$Sr$_x$NiO$_4$ \cite{Anissimova2014}. Understanding the relationship between these four signatures of stripe formation as a function of doping would be a crucial leap forward in our understanding of the cuprates.

Since direct, spectroscopic evidence of fluctuating charge stripes in SC cuprates is lacking, it may be possible to find an avenue of progress through indirect measurements. Recently, it was discovered that the dispersion of the Cu-O bond-stretching longitudinal-optical (LO) phonon in SC \LSCOopt{} (LSCO15, $T_\text{c} = 38\,\text{K}$) \cite{Reznik2007} and \LSCOTwenty{} ($T_\text{c} \approx 35\,\text{K}$) \cite{Park2014} displays a strong anomalous softening interpreted as a coupling to a novel charge collective mode \cite{Park2014}. Furthermore, merely a weak signature of the anomaly is visible in the phonon linewidth of \LSCOseven{} ($T_\text{c} \approx 15\,\text{K}$)  and \LSCOTwentyfive{} ($T_\text{c} \approx 15\,\text{K}$), suggesting that the strength of the anomaly tracks the doping-dependence of $T_c$ \cite{Park2014}.

Similar phonon anomalies have been observed in Bi$_2$Sr$_2$CaCu$_2$O$_{8+\delta}$ \cite{Chaix2017}, \LBCOtwelve{} (LBCO) \cite{Reznik2006}, \LNSCO{} (LNSCO) \cite{Reznik2007} and YBa$_2$Cu$_3$O$_{6.6}$ \cite{LeTacon2014} hinting at a ubiquitous feature of cuprate superconductors.

In order to further investigate the robustness of the phonon anomaly in the $0.1 < x < 0.2$ doping range, we examine two compounds derived from La$_2$CuO$_4$ with unique magnetic, structural and superconducting properties.

\begin{figure}
    \centering
    \includegraphics[width=\columnwidth]{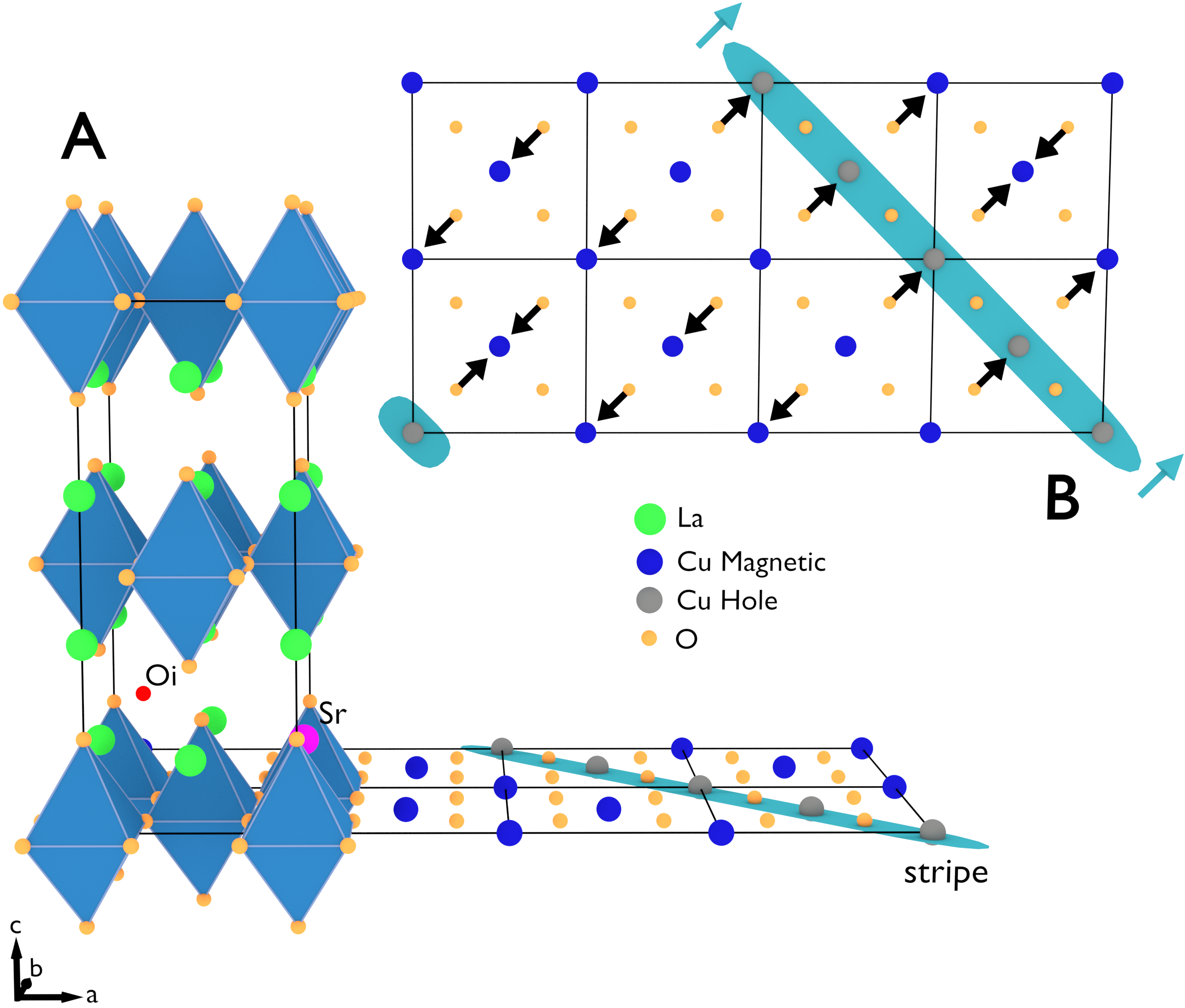}
    \caption{\textbf{A:} Sketch of \LSCOO{} in the orthorhombic unit cell with two distinct dopant species: Sr$^{2+}$ replacing La$^{3+}$ and interstitial oxygen (Oi). The supercell contains 32 formula units, and the two singular dopant species thus represent a doping of $n_h = \frac{3}{32} \approx 0.09$. Blue Cu atoms represent anti-ferromagnetic regions while grey Cu atoms represent a static charge stripe with $\frac{1}{2}$ hole per Cu. \textbf{B:} Sketch of in-phase displacement of dynamic charge domain wall modes coupled to the phonon. Black arrows represent the displacement due to the Cu-O half-breathing mode at $\bm{q}=(0.25,0.25,0)$, while teal arrows represent the in-phase phason mode of the charge stripe \cite{Kaneshita2002}.}
    \label{fig:supercell}
\end{figure}

Hole-doping of the parent compound La$_2$CuO$_4$ can be performed by introducing two distinct dopant species. Replacing La$^{3+}$ for Sr$^{2+}$ yields \LSCO{} (LSCO) with `quenched doping' \cite{Wells1997}, meaning that Sr has a fixed random distribution on La sites after crystal growth. On the other hand, an `annealed doping' \cite{Wells1997} can be obtained by introducing excess, mobile oxygen anions into the lattice by electrochemical methods \cite{Blakeslee1998}, obtaining \LCOO{} (LCO+O). At low Sr concentrations ($x\leq 0.14$) it is possible to combine Sr and O dopants, resulting in `co-doped' \LSCOO{} (LSCO+O) \cite{Mohottala2006}. Fig. \ref{fig:supercell}A depicts the crystallographic sites of O/Sr dopant ions \cite{Radaelli1994,Rial1995,Rial1997a}.

Quenched Sr$^{2+}$ doping creates a superconductor where \Tc{} varies continuously with doping, forming the so-called superconducting dome for $0.05 \leq x \leq 0.25$. Meanwhile the mobile O dopants only seem to allow for certain superconducting phases to emerge ($T_\text{c} = 16, 32, \approx 40\,\text{K}$) due to oxygen content \cite{Liu2005}, pressure \cite{Lorenz2002} or thermal treatment \cite{Fratini2010}.

At sufficient O-doping, LSCO+O is superconducting at ambient pressure below $T_\text{c} \approx 40\,\text{K}$, regardless of Sr-content \cite{Mohottala2006}.  While the underlying mechanism for the connection between annealed disorder and superconductivity is far from settled, the distribution of O dopants is clearly distinct from the distribution of Sr$^{2+}$ dopants. Contrary to the quenched Sr$^{2+}$ dopants, annealed O dopants in LSCO+O are responsible for a number of structural ordering phenomena such as staging \cite{Wells1996, Ray2017} and fractal-like distributions of ordered superstructure patches \cite{Fratini2010, Poccia2012}. 

Differences can also be observed in the magnetic structure of the two compounds. In LSCO+O, the transition to static magnetic stripe order, $T_\text{N}$, roughly coincides with $T_\text{c}$ \cite{Udby2013}. In LSCO, on the other hand, it generally holds that the spins freeze gradually with decreasing temperature and $T_\text{N} <  T_\text{c}$ up to a doping of $x=0.13$, above which the magnetic spectrum becomes gapped \cite{Hirota2001, Julien2003}. This difference in magnetic signatures can partially be explained by the fact that LSCO+O phase separates into distinct spin-stripe ordered ($x \approx \frac{1}{8}$) and optimally superconducting ($x \approx 0.16$) phases, regardless of strontium content prior to oxygenation \cite{Mohottala2006, Udby2013}. 

In this Letter, we investigate the effects of oxygen disorder on the Cu-O bond-stretching phonon mode by providing evidence of phonon anomalies in oxygen-doped LCO+O and co-doped \LSCOOsix{} (LSCO6+O). If the phonon anomaly is related to a specific structural instability due to quenched disorder, one would expect a different response of the lattice in oxygen-doped compounds with annealed disorder. Our results show remarkable similarities of the phonon anomaly in LCO+O, LSCO+O and LSCO. This points towards a purely electronic origin of the phonon anomaly. In order to investigate the possible connection to static stripe order, we explore the effects of a 10 T applied magnetic field. A 10 T field is known to significantly increase the static magnetic response for both LCO+O \cite{Lee2004} and LSCO6+O \cite{Holm2019}. The presence of a field-induced enhancement of the phonon anomaly would indicate a direct connection to static magnetic stripe order, while the absence of a field-effect would point to another source of the anomaly. We observe no detectable change in the phonon signal with applied field, an observation which is compatible with the phonon coupling to {\it dynamic} charge fluctuations as has been suggested previously \cite{Reznik2006}. Our observations thus reinforce and expand this interpretation more broadly for doped cuprates.

The samples are high quality LCO+O and LSCO6+O single crystals. The oxygen-stoichiometric samples were grown by the Traveling Solvent Float Zone method. Subsequently, oxygen intercalation was performed using wet-chemical methods \cite{Blakeslee1998}. Both samples are optimally superconducting with $T_\text{c} = 43\,\text{K}$ (LCO+O) and $T_\text{c} = 37.5\,\text{K}$ (LSCO6+O), confirmed by magnetization measurements (See Fig. S1 in Supplemental Material (SM)). Prior to oxygen intercalation the samples were insulating (LCO) or had $T_\text{c} = 8\,\text{K}$ (LSCO6).

Throughout this Letter, all Miller indices are described with reference to the orthorhombic Bmab space group and the samples are aligned in the $a$-$b$ plane. Lattice parameters were found to be $a=b \approx 5.35(5)\,\text{\AA}$, $c \approx 13.1(1)\,\text{\AA}$ for both samples.

Neutron scattering experiments were performed on the IN8 thermal Triple-Axis Spectrometer at Institut Laue-Langevin, Grenoble \cite{dispersiondata, fielddata}. The instrument was configured with a silicon (311) monochromator and pyrolythic graphite (002) analyser in order to access the desired dynamical range while obtaining the best compromise between neutron intensity and energy resolution ($\approx$ 4-5 meV at 80 meV energy transfer with final wavevector $k_\text{f} = 2.226 \, \text{\AA}^{-1}$). We made use of a position sensitive detector (PSD) with a horizontal spatial resolution of $5\,\text{mm}$ in order to identify and subtract spurious scattering (see Fig. S3 and S4 in SM).

\begin{figure}
    \centering
    \includegraphics[width=\columnwidth]{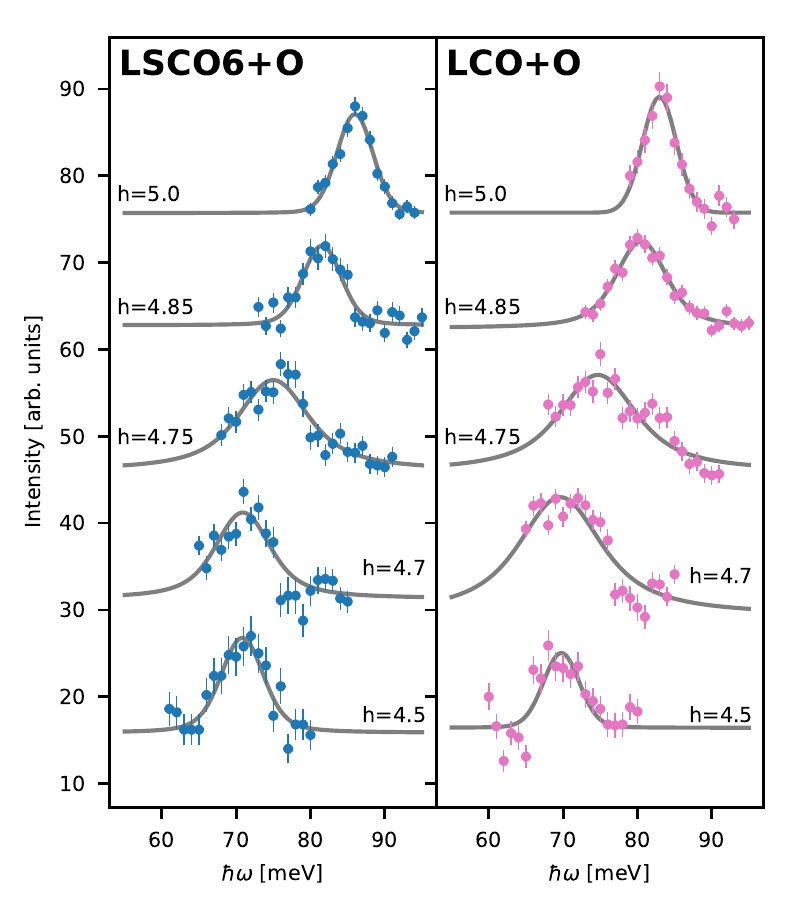}
    \caption{Reduced data at selected wavevectors of the form $Q=(h,h,0)$ for both LSCO6+O and LCO+O at $T=5\,\text{K}$ and $H=0\,\text{T}$. Data at $Q=(5,5,0)$ and $Q=(4.85,4.85,0)$ was scaled by a factor of $\frac{1}{2}$ for clarity due to an increase of intensity from the phonon form factor. Data at different $h$ are offset for clarity. Solid lines are fits to a DHO lineshape (see text).}
    \label{fig:rawdata}
\end{figure}

Figure \ref{fig:rawdata} shows the reduced data at selected wavevectors for both LSCO6+O and LCO+O. The data has been fitted to a Damped Harmonic Oscillator (DHO) model \cite{Fak1997} with a flat background, convoluted with instrument resolution:

\[ S(\bm{q},\omega) = I_\text{ph} \frac{1}{\pi \omega_{\bm{q}}} \frac{\gamma}{(\omega - \omega_{\bm{q}})^2 + \gamma^2} + I_\text{BG} \, , \]

\noindent where $I_\text{ph}$ is the phonon intensity, $\omega_{\bm{q}}$ the phonon energy at wave vector $\bm{q}$, $\gamma$ the phonon linewidth and $I_\text{BG}$ the background intensity.

\begin{figure}
    \centering
    \includegraphics[width=\columnwidth]{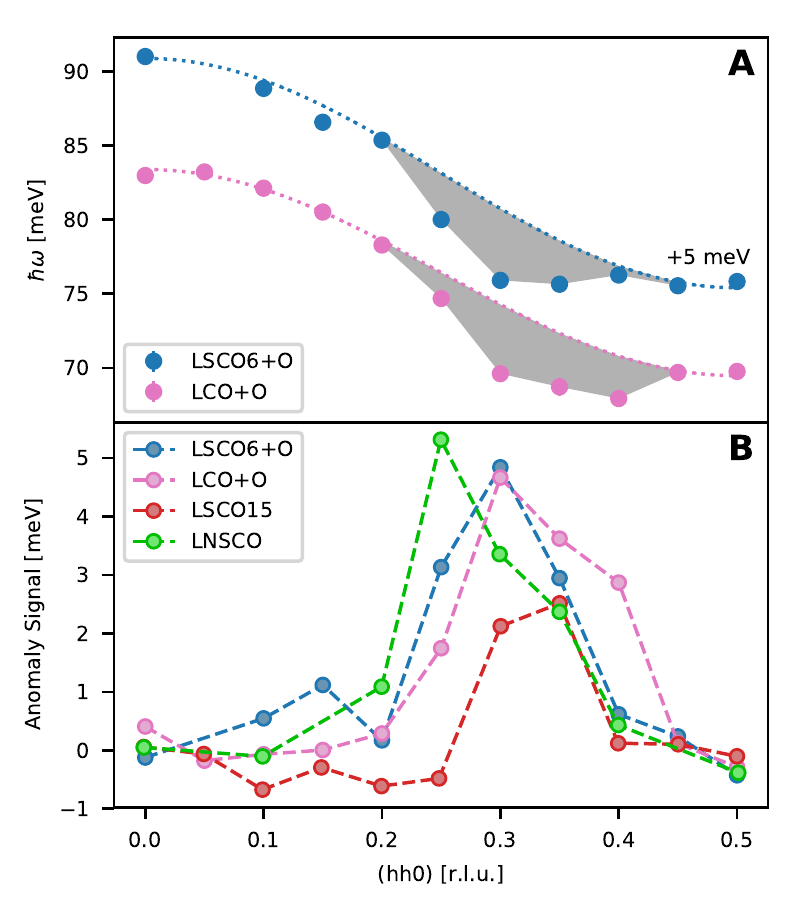}
    \caption{\textbf{(A)}: Dispersion of the LO phonon obtained from the peak positions of individual spectra of both LCO+O and LSCO6+O (offset by $5\,\text{meV}$ for clarity). Error bars smaller than the markers are not shown. Dashed line is the normal sinusoidal dispersion as described in the text. All data was obtained at $T= 5 \, \text{K}$. \textbf{(B)}: Difference between sinusoidal and measured dispersion in \LCOO{} (LCO+O) \LSCOO{} (LSCO+O), \LSCOopt{} (LSCO15) and \LNSCO{} (LNSCO). Data for LSCO15 and LNSCO adapted from \cite{Reznik2007}.}
    \label{fig:dispersion}
\end{figure}

The extracted dispersion from the zero-field data is shown in Fig. \ref{fig:dispersion} along with a normal sinusoidal dispersion, $\hbar\omega_q=\alpha \cos(2 \pi q) + \beta$, inferred from phonon calculations on LSCO using Density Functional Theory (See Fig. S5 in SM and \cite{Giustino2008}). We fit the cosine-function to points near the zone center ($\bm{q}=(0,0,0))$ and edge $(\bm{q}=(\frac{1}{2},\frac{1}{2},0))$ to obtain the dashed curves of Fig. \ref{fig:dispersion}A. To quantify the magnitude of the anomaly, we define the `anomaly signal' as the difference between the normal dispersion and the measured data (gray shaded area in Fig. \ref{fig:dispersion}A). Fig. \ref{fig:dispersion}B shows our anomaly signal for LCO+O and LSCO6+O along with previous results from optimally doped LSCO15 and insulating, stripe-ordered LNSCO \cite{Reznik2007}. We emphasize the presence of similar anomaly signals on an absolute scale across all studied samples. The linewidth broadening $\gamma$ follows the softening of $\omega_{\bm{q}}$ similar to what was observed in LSCO15 and LNSCO \cite{Reznik2007} (see Fig. S2 in SM). Finally, our data can be qualitatively described by a linear combination of stripe ordered and optimally superconducting anomaly signals (See Fig. S7 in SM), consistent with the observation of phase separation in the (super)oxygenated compounds \cite{Mohottala2006}.

To begin the discussion of our results, we remark that softening and/or broadening of phonon modes is generally a signature of an incipient structural or electronic instability. Typical examples include structural phase transitions, $\bm{q}$-dependence of the electron-phonon matrix element, Fermi surface nesting and electronic correlations \cite{Reznik2012}. In order to determine the origin of a given phonon anomaly, it is therefore important to carefully exhaust alternatives before making statements about the connection to novel phases such as dynamic charge stripes.

The phonon anomaly appears in vicinity of the wave vector $\bm{q}_\text{cp}=(\frac{1}{4},\frac{1}{4},0)$, consistent with charge stripe ordering as illustrated in Fig. \ref{fig:supercell}. Measurements of LNSCO \cite{Reznik2008b}, LSCO15 \cite{Reznik2007} and LBCO \cite{Reznik2007} have shown a suppression of the anomaly as one moves away from the bond-stretching direction \cite{Reznik2008b}, supporting a one dimensional stripe-like picture. Since static stripe order has, so far, not been observed in LSCO15 any connection between the phonon anomaly and stripes is likely dynamic. Additionally, the phonon anomaly in LSCO15 and LBCO has almost no temperature dependence apart from a slightly sharper peak shape when heating from $10\,\text{K}$ to $300\,\text{K}$ \cite{Reznik2006,Reznik2007}. These phenomena rule out anharmonicity and structural inhomogeneity as mechanisms for the phonon anomaly in these systems.

A combination of inelastic X-ray and ARPES measurements on overdoped LSCO ($x=0.2$ and $x=0.3$) have shown that the phonon anomaly wavevector is inconsistent with Fermi surface nesting \cite{Park2013,Park2014}, contradicting the idea of a phonon softening due to a Kohn anomaly. A different, possibly $\bm{q}$-dependent, electron-phonon coupling could still be responsible for the phonon anomaly. Such an effect would renormalize the electronic quasiparticle dispersion (the so-called `ARPES kink' \cite{Garcia2010}) at energies similar to the phonon softening. The ARPES kink has been observed in LSCO $x=0.2$ and $x=0.3$, but since only LSCO $x=0.2$ shows anomalous phonons, the two phenomena appear to not be connected \cite{Park2014}.

Thus, all previous studies are unable to explain the phonon anomaly through conventional means and any coupling to stripe order is likely dynamic. One possible scenario is a coupling of the Cu-O bond-stretching phonon with steeply dispersing charge fluctuations. \citeauthor{Kaneshita2002} performed calculations based on the Hubbard model of this scenario, predicting anomalous phonon dispersions due to both transverse (meandering) and longitudinal (compression) coherent stripe fluctuations \cite{Kaneshita2002} (see Fig. \ref{fig:supercell} for a sketch of the transverse mode). We emphasize that the observed phonon anomaly reported here (see Fig. \ref{fig:dispersion}) and in LBCO/LSCO \cite{Reznik2006, Reznik2007} is remarkably similar to the prediction of \citeauthor{Kaneshita2002} (see Fig. 5 in \cite{Kaneshita2002}).

Despite differences in the magnetic excitation spectra as recorded by neutron scattering (including low and zero energy transfers), the three materials LCO+O, LSCO6+O and LSCO15 have remarkably similar in-plane Cu-O bond-stretching dispersions (Fig. \ref{fig:dispersion}). Furthermore, static charge order at zero field has been observed in a different sample of LCO+O \cite{Zhang2018} and in LNSCO \cite{Tranquada1996} but so far not in LSCO6+O nor in LSCO15. These observations together rule out a unique, direct connection between static stripes (spin and charge) and the phonon anomaly. In order to further confirm this point, we performed scans of LSCO6+O at selected wave vectors in a $H=10\,\text{T}$ magnetic field which is known to induce a considerable volume of stripe-like magnetic order in this particular sample \cite{Holm2019}. While static charge order has not been observed in LSCO6+O, measurements on LSCO ($x=0.12$) have shown that static charge and spin stripes respond identically to magnetic fields \cite{Christensen2014}.

\begin{figure}
    \centering
    \includegraphics[width=\columnwidth]{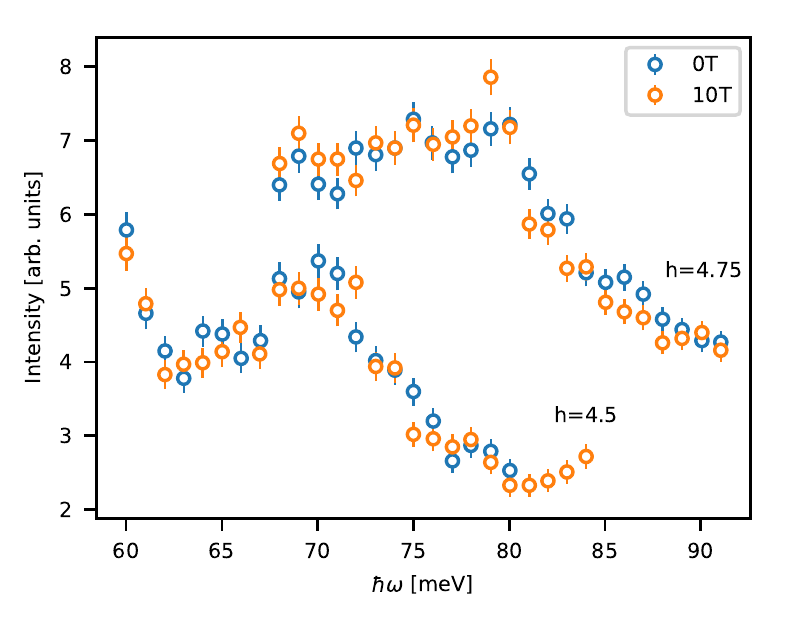}
    \caption{Comparison of representative constant-Q ($h$,$h$,0) scans of \LSCOOsix{} in zero-field and with an applied field of $10\,\text{T}$. All measurements performed at $T=5\,\text{K}$}
    \label{fig:field}
\end{figure}

Figure \ref{fig:field} contains data at two wave vectors with and without an applied magnetic field of $10\,\text{T}$, clearly showing the absence of any detectable field effect on the in-plane Cu-O bond-stretching phonon at $\bm{q}=(\frac{1}{4},\frac{1}{4},0)$. These measurements were performed simultaneously with measurements of the low-energy magnetic fluctuations \cite{Udby2019}, confirming a significant increase in the magnetic spectral weight towards lower energies consistent with the appearance of field-induced stripe-order \cite{Holm2019}. Thus, the appearance of static magnetic stripe order does not affect the phonon anomaly in LSCO6+O. A similar insensitivity of the phonon anomaly to an applied magnetic field has been observed in underdoped ($T_\text{c} = 66\,\text{K}$) YBa$_2$CuO$_{6.6}$ \cite{Reznik2016}.

We have shown that the phonon anomaly is a robust feature in optimally doped  as well as stripe-ordered cuprates which is independent of the structural details related to the doping process. Since it is equally well-formed in stripe-ordered and optimally doped systems, where the latter show no static magnetic order, the anomaly is surprisingly insensitive to low-energy magnetic characteristics. This is further confirmed by the absence of a magnetic field effect in LSCO6+O which introduces static magnetic stripe-order.

The phonon anomaly is strongest in the doping region around optimal $T_\text{c}$ ($0.125 \le n_\text{h} \le 0.20$) (LSCO15, LNSCO \cite{Reznik2007}, LBCO \cite{Reznik2006}, LSCO6+O, LCO+O), regardless of the presence of static charge order (LNSCO \cite{Tranquada1995}, LBCO \cite{Fujita2004}), suppression of bulk superconductivity (LNSCO \cite{Tranquada1996}) or dopant disorder (LCO+O, LSCO6+O). In addition, the phonon anomaly is unaffected by magnetic fields (LSCO6+O, YBa$_2$CuO$_{6.6}$ \cite{Reznik2016}) and temperature (LBCO, LSCO15) \cite{Reznik2007}). Thus it appears to be an intrinsic, robust signature of doped cuprates near optimal doping .

In conclusion, we have measured the in-plane Cu-O bond-stretching phonon in LSCO6+O and LCO+O and provided evidence for significant anomalous behavior. Since one sample (LCO+O) exhibits charge order \cite{Zhang2018} while the other (LSCO6+O) does not \cite{Holm2019}, and since the samples also have distinct magnetic spectra with distinct field dependencies \cite{Jacobsen2018, Holm2019} we conclude that the phonon anomaly has no direct, trivial relationship to either magnetic or charge static order. In addition, the unique structural characteristics of oxygen-doped samples rule out a connection between the specific dopant species and the phonon anomaly. We proceed to conclude that the phonon anomaly is a signature of transverse charge stripe fluctuations. If fluctuating stripes are the fundamental degrees of freedom relevant for the cuprates, it is appealing to draw a connection to Pair-Density-Wave superconductors \cite{Fradkin2015}. In this system, the fundamental degrees of freedom are transverse charge fluctuations in an `electronic liquid crystal' phase without long range order \cite{Kivelson1998}. In this scenario, the phonon anomaly in materials without static stripe order is due to a matching of the phonon wavevector with, otherwise undetectable, short-range transverse stripe correlations. The $x=\frac{1}{8}$ anomaly then corresponds to the special case where stripes exhibit long-range order. 

\bibliography{mainbib}

\begin{thebibliography}{45}%
\makeatletter
\providecommand \@ifxundefined [1]{%
 \@ifx{#1\undefined}
}%
\providecommand \@ifnum [1]{%
 \ifnum #1\expandafter \@firstoftwo
 \else \expandafter \@secondoftwo
 \fi
}%
\providecommand \@ifx [1]{%
 \ifx #1\expandafter \@firstoftwo
 \else \expandafter \@secondoftwo
 \fi
}%
\providecommand \natexlab [1]{#1}%
\providecommand \enquote  [1]{``#1''}%
\providecommand \bibnamefont  [1]{#1}%
\providecommand \bibfnamefont [1]{#1}%
\providecommand \citenamefont [1]{#1}%
\providecommand \href@noop [0]{\@secondoftwo}%
\providecommand \href [0]{\begingroup \@sanitize@url \@href}%
\providecommand \@href[1]{\@@startlink{#1}\@@href}%
\providecommand \@@href[1]{\endgroup#1\@@endlink}%
\providecommand \@sanitize@url [0]{\catcode `\\12\catcode `\$12\catcode
  `\&12\catcode `\#12\catcode `\^12\catcode `\_12\catcode `\%12\relax}%
\providecommand \@@startlink[1]{}%
\providecommand \@@endlink[0]{}%
\providecommand \url  [0]{\begingroup\@sanitize@url \@url }%
\providecommand \@url [1]{\endgroup\@href {#1}{\urlprefix }}%
\providecommand \urlprefix  [0]{URL }%
\providecommand \Eprint [0]{\href }%
\providecommand \doibase [0]{http://dx.doi.org/}%
\providecommand \selectlanguage [0]{\@gobble}%
\providecommand \bibinfo  [0]{\@secondoftwo}%
\providecommand \bibfield  [0]{\@secondoftwo}%
\providecommand \translation [1]{[#1]}%
\providecommand \BibitemOpen [0]{}%
\providecommand \bibitemStop [0]{}%
\providecommand \bibitemNoStop [0]{.\EOS\space}%
\providecommand \EOS [0]{\spacefactor3000\relax}%
\providecommand \BibitemShut  [1]{\csname bibitem#1\endcsname}%
\let\auto@bib@innerbib\@empty
\bibitem [{\citenamefont {Tranquada}\ \emph {et~al.}(1995)\citenamefont
  {Tranquada}, \citenamefont {Sternlieb}, \citenamefont {Axe}, \citenamefont
  {Nakamura},\ and\ \citenamefont {Uchida}}]{Tranquada1995}%
  \BibitemOpen
  \bibfield  {author} {\bibinfo {author} {\bibfnamefont {J.~M.}\ \bibnamefont
  {Tranquada}}, \bibinfo {author} {\bibfnamefont {B.~J.}\ \bibnamefont
  {Sternlieb}}, \bibinfo {author} {\bibfnamefont {J.~D.}\ \bibnamefont {Axe}},
  \bibinfo {author} {\bibfnamefont {Y.}~\bibnamefont {Nakamura}}, \ and\
  \bibinfo {author} {\bibfnamefont {S.}~\bibnamefont {Uchida}},\ }\href
  {\doibase 10.1038/375561a0} {\bibfield  {journal} {\bibinfo  {journal}
  {Nature}\ }\textbf {\bibinfo {volume} {375}},\ \bibinfo {pages} {561}
  (\bibinfo {year} {1995})}\BibitemShut {NoStop}%
\bibitem [{\citenamefont {Tranquada}(2013)}]{Tranquada2013}%
  \BibitemOpen
  \bibfield  {author} {\bibinfo {author} {\bibfnamefont {J.~M.}\ \bibnamefont
  {Tranquada}},\ }\href {\doibase 10.1063/1.4818402} {\bibfield  {journal}
  {\bibinfo  {journal} {AIP Conference Proceedings}\ }\textbf {\bibinfo
  {volume} {1550}},\ \bibinfo {pages} {114} (\bibinfo {year}
  {2013})}\BibitemShut {NoStop}%
\bibitem [{\citenamefont {Tranquada}\ \emph {et~al.}(1996)\citenamefont
  {Tranquada}, \citenamefont {Axe}, \citenamefont {Ichikawa}, \citenamefont
  {Nakamura}, \citenamefont {Uchida},\ and\ \citenamefont
  {Nachumi}}]{Tranquada1996}%
  \BibitemOpen
  \bibfield  {author} {\bibinfo {author} {\bibfnamefont {J.~M.}\ \bibnamefont
  {Tranquada}}, \bibinfo {author} {\bibfnamefont {J.~D.}\ \bibnamefont {Axe}},
  \bibinfo {author} {\bibfnamefont {N.}~\bibnamefont {Ichikawa}}, \bibinfo
  {author} {\bibfnamefont {Y.}~\bibnamefont {Nakamura}}, \bibinfo {author}
  {\bibfnamefont {S.}~\bibnamefont {Uchida}}, \ and\ \bibinfo {author}
  {\bibfnamefont {B.}~\bibnamefont {Nachumi}},\ }\href {\doibase
  10.1103/PhysRevB.54.7489} {\bibfield  {journal} {\bibinfo  {journal} {Phys.
  Rev. B}\ }\textbf {\bibinfo {volume} {54}},\ \bibinfo {pages} {7489}
  (\bibinfo {year} {1996})}\BibitemShut {NoStop}%
\bibitem [{\citenamefont {Christensen}\ \emph {et~al.}(2014)\citenamefont
  {Christensen}, \citenamefont {Chang}, \citenamefont {Larsen}, \citenamefont
  {Fujita}, \citenamefont {Oda}, \citenamefont {Ido}, \citenamefont {Momono},
  \citenamefont {Forgan}, \citenamefont {Holmes}, \citenamefont {Mesot},
  \citenamefont {Huecker},\ and\ \citenamefont
  {v~Zimmermann}}]{Christensen2014}%
  \BibitemOpen
  \bibfield  {author} {\bibinfo {author} {\bibfnamefont {N.~B.}\ \bibnamefont
  {Christensen}}, \bibinfo {author} {\bibfnamefont {J.}~\bibnamefont {Chang}},
  \bibinfo {author} {\bibfnamefont {J.}~\bibnamefont {Larsen}}, \bibinfo
  {author} {\bibfnamefont {M.}~\bibnamefont {Fujita}}, \bibinfo {author}
  {\bibfnamefont {M.}~\bibnamefont {Oda}}, \bibinfo {author} {\bibfnamefont
  {M.}~\bibnamefont {Ido}}, \bibinfo {author} {\bibfnamefont {N.}~\bibnamefont
  {Momono}}, \bibinfo {author} {\bibfnamefont {E.~M.}\ \bibnamefont {Forgan}},
  \bibinfo {author} {\bibfnamefont {A.~T.}\ \bibnamefont {Holmes}}, \bibinfo
  {author} {\bibfnamefont {J.}~\bibnamefont {Mesot}}, \bibinfo {author}
  {\bibfnamefont {M.}~\bibnamefont {Huecker}}, \ and\ \bibinfo {author}
  {\bibfnamefont {M.}~\bibnamefont {v~Zimmermann}},\ }\href@noop {} {\bibfield
  {journal} {\bibinfo  {journal} {arXiv:1404.3192 [cond-mat]}\ } (\bibinfo
  {year} {2014})},\ \Eprint {http://arxiv.org/abs/1404.3192} {arXiv:1404.3192
  [cond-mat]} \BibitemShut {NoStop}%
\bibitem [{\citenamefont {Thampy}\ \emph {et~al.}(2014)\citenamefont {Thampy},
  \citenamefont {Dean}, \citenamefont {Christensen}, \citenamefont {Steinke},
  \citenamefont {Islam}, \citenamefont {Oda}, \citenamefont {Ido},
  \citenamefont {Momono}, \citenamefont {Wilkins},\ and\ \citenamefont
  {Hill}}]{Thampy2014}%
  \BibitemOpen
  \bibfield  {author} {\bibinfo {author} {\bibfnamefont {V.}~\bibnamefont
  {Thampy}}, \bibinfo {author} {\bibfnamefont {M.~P.~M.}\ \bibnamefont {Dean}},
  \bibinfo {author} {\bibfnamefont {N.~B.}\ \bibnamefont {Christensen}},
  \bibinfo {author} {\bibfnamefont {L.}~\bibnamefont {Steinke}}, \bibinfo
  {author} {\bibfnamefont {Z.}~\bibnamefont {Islam}}, \bibinfo {author}
  {\bibfnamefont {M.}~\bibnamefont {Oda}}, \bibinfo {author} {\bibfnamefont
  {M.}~\bibnamefont {Ido}}, \bibinfo {author} {\bibfnamefont {N.}~\bibnamefont
  {Momono}}, \bibinfo {author} {\bibfnamefont {S.~B.}\ \bibnamefont {Wilkins}},
  \ and\ \bibinfo {author} {\bibfnamefont {J.~P.}\ \bibnamefont {Hill}},\
  }\href {\doibase 10.1103/PhysRevB.90.100510} {\bibfield  {journal} {\bibinfo
  {journal} {Phys. Rev. B}\ }\textbf {\bibinfo {volume} {90}},\ \bibinfo
  {pages} {100510} (\bibinfo {year} {2014})}\BibitemShut {NoStop}%
\bibitem [{\citenamefont {Croft}\ \emph {et~al.}(2014)\citenamefont {Croft},
  \citenamefont {Lester}, \citenamefont {Senn}, \citenamefont {Bombardi},\ and\
  \citenamefont {Hayden}}]{Croft2014}%
  \BibitemOpen
  \bibfield  {author} {\bibinfo {author} {\bibfnamefont {T.~P.}\ \bibnamefont
  {Croft}}, \bibinfo {author} {\bibfnamefont {C.}~\bibnamefont {Lester}},
  \bibinfo {author} {\bibfnamefont {M.~S.}\ \bibnamefont {Senn}}, \bibinfo
  {author} {\bibfnamefont {A.}~\bibnamefont {Bombardi}}, \ and\ \bibinfo
  {author} {\bibfnamefont {S.~M.}\ \bibnamefont {Hayden}},\ }\href {\doibase
  10.1103/PhysRevB.89.224513} {\bibfield  {journal} {\bibinfo  {journal} {Phys.
  Rev. B}\ }\textbf {\bibinfo {volume} {89}},\ \bibinfo {pages} {224513}
  (\bibinfo {year} {2014})}\BibitemShut {NoStop}%
\bibitem [{\citenamefont {Anissimova}\ \emph {et~al.}(2014)\citenamefont
  {Anissimova}, \citenamefont {Parshall}, \citenamefont {Gu}, \citenamefont
  {Marty}, \citenamefont {Lumsden}, \citenamefont {Chi}, \citenamefont
  {{Fernandez-Baca}}, \citenamefont {Abernathy}, \citenamefont {Lamago},
  \citenamefont {Tranquada},\ and\ \citenamefont {Reznik}}]{Anissimova2014}%
  \BibitemOpen
  \bibfield  {author} {\bibinfo {author} {\bibfnamefont {S.}~\bibnamefont
  {Anissimova}}, \bibinfo {author} {\bibfnamefont {D.}~\bibnamefont
  {Parshall}}, \bibinfo {author} {\bibfnamefont {G.~D.}\ \bibnamefont {Gu}},
  \bibinfo {author} {\bibfnamefont {K.}~\bibnamefont {Marty}}, \bibinfo
  {author} {\bibfnamefont {M.~D.}\ \bibnamefont {Lumsden}}, \bibinfo {author}
  {\bibfnamefont {S.}~\bibnamefont {Chi}}, \bibinfo {author} {\bibfnamefont
  {J.~A.}\ \bibnamefont {{Fernandez-Baca}}}, \bibinfo {author} {\bibfnamefont
  {D.~L.}\ \bibnamefont {Abernathy}}, \bibinfo {author} {\bibfnamefont
  {D.}~\bibnamefont {Lamago}}, \bibinfo {author} {\bibfnamefont {J.~M.}\
  \bibnamefont {Tranquada}}, \ and\ \bibinfo {author} {\bibfnamefont
  {D.}~\bibnamefont {Reznik}},\ }\href {\doibase 10.1038/ncomms4467} {\bibfield
   {journal} {\bibinfo  {journal} {Nature Communications}\ }\textbf {\bibinfo
  {volume} {5}},\ \bibinfo {pages} {3467} (\bibinfo {year} {2014})}\BibitemShut
  {NoStop}%
\bibitem [{\citenamefont {Reznik}\ \emph {et~al.}(2007)\citenamefont {Reznik},
  \citenamefont {Pintschovius}, \citenamefont {Fujita}, \citenamefont {Yamada},
  \citenamefont {Gu},\ and\ \citenamefont {Tranquada}}]{Reznik2007}%
  \BibitemOpen
  \bibfield  {author} {\bibinfo {author} {\bibfnamefont {D.}~\bibnamefont
  {Reznik}}, \bibinfo {author} {\bibfnamefont {L.}~\bibnamefont
  {Pintschovius}}, \bibinfo {author} {\bibfnamefont {M.}~\bibnamefont
  {Fujita}}, \bibinfo {author} {\bibfnamefont {K.}~\bibnamefont {Yamada}},
  \bibinfo {author} {\bibfnamefont {G.~D.}\ \bibnamefont {Gu}}, \ and\ \bibinfo
  {author} {\bibfnamefont {J.~M.}\ \bibnamefont {Tranquada}},\ }\href {\doibase
  10.1007/s10909-007-9318-9} {\bibfield  {journal} {\bibinfo  {journal} {J Low
  Temp Phys}\ }\textbf {\bibinfo {volume} {147}},\ \bibinfo {pages} {353}
  (\bibinfo {year} {2007})}\BibitemShut {NoStop}%
\bibitem [{\citenamefont {Park}\ \emph {et~al.}(2014)\citenamefont {Park},
  \citenamefont {Fukuda}, \citenamefont {Hamann}, \citenamefont {Lamago},
  \citenamefont {Pintschovius}, \citenamefont {Fujita}, \citenamefont
  {Yamada},\ and\ \citenamefont {Reznik}}]{Park2014}%
  \BibitemOpen
  \bibfield  {author} {\bibinfo {author} {\bibfnamefont {S.~R.}\ \bibnamefont
  {Park}}, \bibinfo {author} {\bibfnamefont {T.}~\bibnamefont {Fukuda}},
  \bibinfo {author} {\bibfnamefont {A.}~\bibnamefont {Hamann}}, \bibinfo
  {author} {\bibfnamefont {D.}~\bibnamefont {Lamago}}, \bibinfo {author}
  {\bibfnamefont {L.}~\bibnamefont {Pintschovius}}, \bibinfo {author}
  {\bibfnamefont {M.}~\bibnamefont {Fujita}}, \bibinfo {author} {\bibfnamefont
  {K.}~\bibnamefont {Yamada}}, \ and\ \bibinfo {author} {\bibfnamefont
  {D.}~\bibnamefont {Reznik}},\ }\href {\doibase 10.1103/PhysRevB.89.020506}
  {\bibfield  {journal} {\bibinfo  {journal} {Phys. Rev. B}\ }\textbf {\bibinfo
  {volume} {89}},\ \bibinfo {pages} {020506} (\bibinfo {year}
  {2014})}\BibitemShut {NoStop}%
\bibitem [{\citenamefont {Chaix}\ \emph {et~al.}(2017)\citenamefont {Chaix},
  \citenamefont {Ghiringhelli}, \citenamefont {Peng}, \citenamefont
  {Hashimoto}, \citenamefont {Moritz}, \citenamefont {Kummer}, \citenamefont
  {Brookes}, \citenamefont {He}, \citenamefont {Chen}, \citenamefont {Ishida},
  \citenamefont {Yoshida}, \citenamefont {Eisaki}, \citenamefont {Salluzzo},
  \citenamefont {Braicovich}, \citenamefont {Shen}, \citenamefont {Devereaux},\
  and\ \citenamefont {Lee}}]{Chaix2017}%
  \BibitemOpen
  \bibfield  {author} {\bibinfo {author} {\bibfnamefont {L.}~\bibnamefont
  {Chaix}}, \bibinfo {author} {\bibfnamefont {G.}~\bibnamefont {Ghiringhelli}},
  \bibinfo {author} {\bibfnamefont {Y.~Y.}\ \bibnamefont {Peng}}, \bibinfo
  {author} {\bibfnamefont {M.}~\bibnamefont {Hashimoto}}, \bibinfo {author}
  {\bibfnamefont {B.}~\bibnamefont {Moritz}}, \bibinfo {author} {\bibfnamefont
  {K.}~\bibnamefont {Kummer}}, \bibinfo {author} {\bibfnamefont {N.~B.}\
  \bibnamefont {Brookes}}, \bibinfo {author} {\bibfnamefont {Y.}~\bibnamefont
  {He}}, \bibinfo {author} {\bibfnamefont {S.}~\bibnamefont {Chen}}, \bibinfo
  {author} {\bibfnamefont {S.}~\bibnamefont {Ishida}}, \bibinfo {author}
  {\bibfnamefont {Y.}~\bibnamefont {Yoshida}}, \bibinfo {author} {\bibfnamefont
  {H.}~\bibnamefont {Eisaki}}, \bibinfo {author} {\bibfnamefont
  {M.}~\bibnamefont {Salluzzo}}, \bibinfo {author} {\bibfnamefont
  {L.}~\bibnamefont {Braicovich}}, \bibinfo {author} {\bibfnamefont {Z.-X.}\
  \bibnamefont {Shen}}, \bibinfo {author} {\bibfnamefont {T.~P.}\ \bibnamefont
  {Devereaux}}, \ and\ \bibinfo {author} {\bibfnamefont {W.-S.}\ \bibnamefont
  {Lee}},\ }\href {\doibase 10.1038/nphys4157} {\bibfield  {journal} {\bibinfo
  {journal} {Nature Physics}\ }\textbf {\bibinfo {volume} {13}},\ \bibinfo
  {pages} {952} (\bibinfo {year} {2017})}\BibitemShut {NoStop}%
\bibitem [{\citenamefont {Reznik}\ \emph {et~al.}(2006)\citenamefont {Reznik},
  \citenamefont {Pintschovius}, \citenamefont {Ito}, \citenamefont {Iikubo},
  \citenamefont {Sato}, \citenamefont {Goka}, \citenamefont {Fujita},
  \citenamefont {Yamada}, \citenamefont {Gu},\ and\ \citenamefont
  {Tranquada}}]{Reznik2006}%
  \BibitemOpen
  \bibfield  {author} {\bibinfo {author} {\bibfnamefont {D.}~\bibnamefont
  {Reznik}}, \bibinfo {author} {\bibfnamefont {L.}~\bibnamefont
  {Pintschovius}}, \bibinfo {author} {\bibfnamefont {M.}~\bibnamefont {Ito}},
  \bibinfo {author} {\bibfnamefont {S.}~\bibnamefont {Iikubo}}, \bibinfo
  {author} {\bibfnamefont {M.}~\bibnamefont {Sato}}, \bibinfo {author}
  {\bibfnamefont {H.}~\bibnamefont {Goka}}, \bibinfo {author} {\bibfnamefont
  {M.}~\bibnamefont {Fujita}}, \bibinfo {author} {\bibfnamefont
  {K.}~\bibnamefont {Yamada}}, \bibinfo {author} {\bibfnamefont {G.~D.}\
  \bibnamefont {Gu}}, \ and\ \bibinfo {author} {\bibfnamefont {J.~M.}\
  \bibnamefont {Tranquada}},\ }\href {\doibase 10.1038/nature04704} {\bibfield
  {journal} {\bibinfo  {journal} {Nature}\ }\textbf {\bibinfo {volume} {440}},\
  \bibinfo {pages} {1170} (\bibinfo {year} {2006})}\BibitemShut {NoStop}%
\bibitem [{\citenamefont {Le~Tacon}\ \emph {et~al.}(2014)\citenamefont
  {Le~Tacon}, \citenamefont {Bosak}, \citenamefont {Souliou}, \citenamefont
  {Dellea}, \citenamefont {Loew}, \citenamefont {Heid}, \citenamefont {Bohnen},
  \citenamefont {Ghiringhelli}, \citenamefont {Krisch},\ and\ \citenamefont
  {Keimer}}]{LeTacon2014}%
  \BibitemOpen
  \bibfield  {author} {\bibinfo {author} {\bibfnamefont {M.}~\bibnamefont
  {Le~Tacon}}, \bibinfo {author} {\bibfnamefont {A.}~\bibnamefont {Bosak}},
  \bibinfo {author} {\bibfnamefont {S.~M.}\ \bibnamefont {Souliou}}, \bibinfo
  {author} {\bibfnamefont {G.}~\bibnamefont {Dellea}}, \bibinfo {author}
  {\bibfnamefont {T.}~\bibnamefont {Loew}}, \bibinfo {author} {\bibfnamefont
  {R.}~\bibnamefont {Heid}}, \bibinfo {author} {\bibfnamefont {K.-P.}\
  \bibnamefont {Bohnen}}, \bibinfo {author} {\bibfnamefont {G.}~\bibnamefont
  {Ghiringhelli}}, \bibinfo {author} {\bibfnamefont {M.}~\bibnamefont
  {Krisch}}, \ and\ \bibinfo {author} {\bibfnamefont {B.}~\bibnamefont
  {Keimer}},\ }\href {\doibase 10.1038/nphys2805} {\bibfield  {journal}
  {\bibinfo  {journal} {Nature Physics}\ }\textbf {\bibinfo {volume} {10}},\
  \bibinfo {pages} {52} (\bibinfo {year} {2014})}\BibitemShut {NoStop}%
\bibitem [{\citenamefont {Wells}\ \emph {et~al.}(1997)\citenamefont {Wells},
  \citenamefont {Lee}, \citenamefont {Kastner}, \citenamefont {Christianson},
  \citenamefont {Birgeneau}, \citenamefont {Yamada}, \citenamefont {Endoh},\
  and\ \citenamefont {Shirane}}]{Wells1997}%
  \BibitemOpen
  \bibfield  {author} {\bibinfo {author} {\bibfnamefont {B.~O.}\ \bibnamefont
  {Wells}}, \bibinfo {author} {\bibfnamefont {Y.~S.}\ \bibnamefont {Lee}},
  \bibinfo {author} {\bibfnamefont {M.~A.}\ \bibnamefont {Kastner}}, \bibinfo
  {author} {\bibfnamefont {R.~J.}\ \bibnamefont {Christianson}}, \bibinfo
  {author} {\bibfnamefont {R.~J.}\ \bibnamefont {Birgeneau}}, \bibinfo {author}
  {\bibfnamefont {K.}~\bibnamefont {Yamada}}, \bibinfo {author} {\bibfnamefont
  {Y.}~\bibnamefont {Endoh}}, \ and\ \bibinfo {author} {\bibfnamefont
  {G.}~\bibnamefont {Shirane}},\ }\href {\doibase
  10.1126/science.277.5329.1067} {\bibfield  {journal} {\bibinfo  {journal}
  {Science}\ }\textbf {\bibinfo {volume} {277}},\ \bibinfo {pages} {1067}
  (\bibinfo {year} {1997})}\BibitemShut {NoStop}%
\bibitem [{\citenamefont {Blakeslee}\ \emph {et~al.}(1998)\citenamefont
  {Blakeslee}, \citenamefont {Birgeneau}, \citenamefont {Chou}, \citenamefont
  {Christianson}, \citenamefont {Kastner}, \citenamefont {Lee},\ and\
  \citenamefont {Wells}}]{Blakeslee1998}%
  \BibitemOpen
  \bibfield  {author} {\bibinfo {author} {\bibfnamefont {P.}~\bibnamefont
  {Blakeslee}}, \bibinfo {author} {\bibfnamefont {R.~J.}\ \bibnamefont
  {Birgeneau}}, \bibinfo {author} {\bibfnamefont {F.~C.}\ \bibnamefont {Chou}},
  \bibinfo {author} {\bibfnamefont {R.}~\bibnamefont {Christianson}}, \bibinfo
  {author} {\bibfnamefont {M.~A.}\ \bibnamefont {Kastner}}, \bibinfo {author}
  {\bibfnamefont {Y.~S.}\ \bibnamefont {Lee}}, \ and\ \bibinfo {author}
  {\bibfnamefont {B.~O.}\ \bibnamefont {Wells}},\ }\href {\doibase
  10.1103/PhysRevB.57.13915} {\bibfield  {journal} {\bibinfo  {journal} {Phys.
  Rev. B}\ }\textbf {\bibinfo {volume} {57}},\ \bibinfo {pages} {13915}
  (\bibinfo {year} {1998})}\BibitemShut {NoStop}%
\bibitem [{\citenamefont {Mohottala}\ \emph {et~al.}(2006)\citenamefont
  {Mohottala}, \citenamefont {Wells}, \citenamefont {Budnick}, \citenamefont
  {Hines}, \citenamefont {Niedermayer}, \citenamefont {Udby}, \citenamefont
  {Bernhard}, \citenamefont {Moodenbaugh},\ and\ \citenamefont
  {Chou}}]{Mohottala2006}%
  \BibitemOpen
  \bibfield  {author} {\bibinfo {author} {\bibfnamefont {H.~E.}\ \bibnamefont
  {Mohottala}}, \bibinfo {author} {\bibfnamefont {B.~O.}\ \bibnamefont
  {Wells}}, \bibinfo {author} {\bibfnamefont {J.~I.}\ \bibnamefont {Budnick}},
  \bibinfo {author} {\bibfnamefont {W.~A.}\ \bibnamefont {Hines}}, \bibinfo
  {author} {\bibfnamefont {C.}~\bibnamefont {Niedermayer}}, \bibinfo {author}
  {\bibfnamefont {L.}~\bibnamefont {Udby}}, \bibinfo {author} {\bibfnamefont
  {C.}~\bibnamefont {Bernhard}}, \bibinfo {author} {\bibfnamefont {A.~R.}\
  \bibnamefont {Moodenbaugh}}, \ and\ \bibinfo {author} {\bibfnamefont {F.-C.}\
  \bibnamefont {Chou}},\ }\href {\doibase 10.1038/nmat1633} {\bibfield
  {journal} {\bibinfo  {journal} {Nature Materials}\ }\textbf {\bibinfo
  {volume} {5}},\ \bibinfo {pages} {377} (\bibinfo {year} {2006})}\BibitemShut
  {NoStop}%
\bibitem [{\citenamefont {Radaelli}\ \emph {et~al.}(1994)\citenamefont
  {Radaelli}, \citenamefont {Hinks}, \citenamefont {Mitchell}, \citenamefont
  {Hunter}, \citenamefont {Wagner}, \citenamefont {Dabrowski}, \citenamefont
  {Vandervoort}, \citenamefont {Viswanathan},\ and\ \citenamefont
  {Jorgensen}}]{Radaelli1994}%
  \BibitemOpen
  \bibfield  {author} {\bibinfo {author} {\bibfnamefont {P.~G.}\ \bibnamefont
  {Radaelli}}, \bibinfo {author} {\bibfnamefont {D.~G.}\ \bibnamefont {Hinks}},
  \bibinfo {author} {\bibfnamefont {A.~W.}\ \bibnamefont {Mitchell}}, \bibinfo
  {author} {\bibfnamefont {B.~A.}\ \bibnamefont {Hunter}}, \bibinfo {author}
  {\bibfnamefont {J.~L.}\ \bibnamefont {Wagner}}, \bibinfo {author}
  {\bibfnamefont {B.}~\bibnamefont {Dabrowski}}, \bibinfo {author}
  {\bibfnamefont {K.~G.}\ \bibnamefont {Vandervoort}}, \bibinfo {author}
  {\bibfnamefont {H.~K.}\ \bibnamefont {Viswanathan}}, \ and\ \bibinfo {author}
  {\bibfnamefont {J.~D.}\ \bibnamefont {Jorgensen}},\ }\href {\doibase
  10.1103/PhysRevB.49.4163} {\bibfield  {journal} {\bibinfo  {journal} {Phys.
  Rev. B}\ }\textbf {\bibinfo {volume} {49}},\ \bibinfo {pages} {4163}
  (\bibinfo {year} {1994})}\BibitemShut {NoStop}%
\bibitem [{\citenamefont {Rial}\ \emph {et~al.}(1995)\citenamefont {Rial},
  \citenamefont {Mor\'an}, \citenamefont {{Alario-Franco}}, \citenamefont
  {Amador},\ and\ \citenamefont {Andersen}}]{Rial1995}%
  \BibitemOpen
  \bibfield  {author} {\bibinfo {author} {\bibfnamefont {C.}~\bibnamefont
  {Rial}}, \bibinfo {author} {\bibfnamefont {E.}~\bibnamefont {Mor\'an}},
  \bibinfo {author} {\bibfnamefont {M.~A.}\ \bibnamefont {{Alario-Franco}}},
  \bibinfo {author} {\bibfnamefont {U.}~\bibnamefont {Amador}}, \ and\ \bibinfo
  {author} {\bibfnamefont {N.~H.}\ \bibnamefont {Andersen}},\ }\href {\doibase
  10.1016/0921-4534(95)00570-6} {\bibfield  {journal} {\bibinfo  {journal}
  {Physica C: Superconductivity}\ }\textbf {\bibinfo {volume} {254}},\ \bibinfo
  {pages} {233} (\bibinfo {year} {1995})}\BibitemShut {NoStop}%
\bibitem [{\citenamefont {Rial}\ \emph {et~al.}(1997)\citenamefont {Rial},
  \citenamefont {Mor\'an}, \citenamefont {{Alario-Franco}}, \citenamefont
  {Amador},\ and\ \citenamefont {Andersen}}]{Rial1997a}%
  \BibitemOpen
  \bibfield  {author} {\bibinfo {author} {\bibfnamefont {C.}~\bibnamefont
  {Rial}}, \bibinfo {author} {\bibfnamefont {E.}~\bibnamefont {Mor\'an}},
  \bibinfo {author} {\bibfnamefont {M.~A.}\ \bibnamefont {{Alario-Franco}}},
  \bibinfo {author} {\bibfnamefont {U.}~\bibnamefont {Amador}}, \ and\ \bibinfo
  {author} {\bibfnamefont {N.~H.}\ \bibnamefont {Andersen}},\ }\href {\doibase
  10.1016/S0921-4534(97)00097-X} {\bibfield  {journal} {\bibinfo  {journal}
  {Physica C: Superconductivity}\ }\textbf {\bibinfo {volume} {278}},\ \bibinfo
  {pages} {122} (\bibinfo {year} {1997})}\BibitemShut {NoStop}%
\bibitem [{\citenamefont {Liu}\ \emph {et~al.}(2005)\citenamefont {Liu},
  \citenamefont {Che}, \citenamefont {Zhao},\ and\ \citenamefont
  {Zhao}}]{Liu2005}%
  \BibitemOpen
  \bibfield  {author} {\bibinfo {author} {\bibfnamefont {L.~H.}\ \bibnamefont
  {Liu}}, \bibinfo {author} {\bibfnamefont {G.~C.}\ \bibnamefont {Che}},
  \bibinfo {author} {\bibfnamefont {J.}~\bibnamefont {Zhao}}, \ and\ \bibinfo
  {author} {\bibfnamefont {Z.~X.}\ \bibnamefont {Zhao}},\ }\href {\doibase
  10.1016/j.physc.2005.05.014} {\bibfield  {journal} {\bibinfo  {journal}
  {Physica C: Superconductivity}\ }\textbf {\bibinfo {volume} {425}},\ \bibinfo
  {pages} {37} (\bibinfo {year} {2005})}\BibitemShut {NoStop}%
\bibitem [{\citenamefont {Lorenz}\ \emph {et~al.}(2002)\citenamefont {Lorenz},
  \citenamefont {Li}, \citenamefont {Honma},\ and\ \citenamefont
  {Hor}}]{Lorenz2002}%
  \BibitemOpen
  \bibfield  {author} {\bibinfo {author} {\bibfnamefont {B.}~\bibnamefont
  {Lorenz}}, \bibinfo {author} {\bibfnamefont {Z.~G.}\ \bibnamefont {Li}},
  \bibinfo {author} {\bibfnamefont {T.}~\bibnamefont {Honma}}, \ and\ \bibinfo
  {author} {\bibfnamefont {P.-H.}\ \bibnamefont {Hor}},\ }\href {\doibase
  10.1103/PhysRevB.65.144522} {\bibfield  {journal} {\bibinfo  {journal} {Phys.
  Rev. B}\ }\textbf {\bibinfo {volume} {65}},\ \bibinfo {pages} {144522}
  (\bibinfo {year} {2002})}\BibitemShut {NoStop}%
\bibitem [{\citenamefont {Fratini}\ \emph {et~al.}(2010)\citenamefont
  {Fratini}, \citenamefont {Poccia}, \citenamefont {Ricci}, \citenamefont
  {Campi}, \citenamefont {Burghammer}, \citenamefont {Aeppli},\ and\
  \citenamefont {Bianconi}}]{Fratini2010}%
  \BibitemOpen
  \bibfield  {author} {\bibinfo {author} {\bibfnamefont {M.}~\bibnamefont
  {Fratini}}, \bibinfo {author} {\bibfnamefont {N.}~\bibnamefont {Poccia}},
  \bibinfo {author} {\bibfnamefont {A.}~\bibnamefont {Ricci}}, \bibinfo
  {author} {\bibfnamefont {G.}~\bibnamefont {Campi}}, \bibinfo {author}
  {\bibfnamefont {M.}~\bibnamefont {Burghammer}}, \bibinfo {author}
  {\bibfnamefont {G.}~\bibnamefont {Aeppli}}, \ and\ \bibinfo {author}
  {\bibfnamefont {A.}~\bibnamefont {Bianconi}},\ }\href {\doibase
  10.1038/nature09260} {\bibfield  {journal} {\bibinfo  {journal} {Nature}\
  }\textbf {\bibinfo {volume} {466}},\ \bibinfo {pages} {841} (\bibinfo {year}
  {2010})}\BibitemShut {NoStop}%
\bibitem [{\citenamefont {Wells}\ \emph {et~al.}(1996)\citenamefont {Wells},
  \citenamefont {Birgeneau}, \citenamefont {Chou}, \citenamefont {Endoh},
  \citenamefont {Johnston}, \citenamefont {Kastner}, \citenamefont {Lee},
  \citenamefont {Shirane}, \citenamefont {Tranquada},\ and\ \citenamefont
  {Yamada}}]{Wells1996}%
  \BibitemOpen
  \bibfield  {author} {\bibinfo {author} {\bibfnamefont {B.~O.}\ \bibnamefont
  {Wells}}, \bibinfo {author} {\bibfnamefont {R.~J.}\ \bibnamefont
  {Birgeneau}}, \bibinfo {author} {\bibfnamefont {F.~C.}\ \bibnamefont {Chou}},
  \bibinfo {author} {\bibfnamefont {Y.}~\bibnamefont {Endoh}}, \bibinfo
  {author} {\bibfnamefont {D.~C.}\ \bibnamefont {Johnston}}, \bibinfo {author}
  {\bibfnamefont {M.~A.}\ \bibnamefont {Kastner}}, \bibinfo {author}
  {\bibfnamefont {Y.~S.}\ \bibnamefont {Lee}}, \bibinfo {author} {\bibfnamefont
  {G.}~\bibnamefont {Shirane}}, \bibinfo {author} {\bibfnamefont {J.~M.}\
  \bibnamefont {Tranquada}}, \ and\ \bibinfo {author} {\bibfnamefont
  {K.}~\bibnamefont {Yamada}},\ }\href {\doibase 10.1007/s002570050158}
  {\bibfield  {journal} {\bibinfo  {journal} {Zeitschrift f\"ur Physik B
  Condensed Matter}\ }\textbf {\bibinfo {volume} {100}},\ \bibinfo {pages}
  {535} (\bibinfo {year} {1996})}\BibitemShut {NoStop}%
\bibitem [{\citenamefont {Ray}\ \emph {et~al.}(2017)\citenamefont {Ray},
  \citenamefont {Andersen}, \citenamefont {Jensen}, \citenamefont {Mohottala},
  \citenamefont {Niedermayer}, \citenamefont {Lefmann}, \citenamefont {Wells},
  \citenamefont {v.~Zimmermann},\ and\ \citenamefont {Udby}}]{Ray2017}%
  \BibitemOpen
  \bibfield  {author} {\bibinfo {author} {\bibfnamefont {P.~J.}\ \bibnamefont
  {Ray}}, \bibinfo {author} {\bibfnamefont {N.~H.}\ \bibnamefont {Andersen}},
  \bibinfo {author} {\bibfnamefont {T.~B.~S.}\ \bibnamefont {Jensen}}, \bibinfo
  {author} {\bibfnamefont {H.~E.}\ \bibnamefont {Mohottala}}, \bibinfo {author}
  {\bibfnamefont {C.}~\bibnamefont {Niedermayer}}, \bibinfo {author}
  {\bibfnamefont {K.}~\bibnamefont {Lefmann}}, \bibinfo {author} {\bibfnamefont
  {B.~O.}\ \bibnamefont {Wells}}, \bibinfo {author} {\bibfnamefont
  {M.}~\bibnamefont {v.~Zimmermann}}, \ and\ \bibinfo {author} {\bibfnamefont
  {L.}~\bibnamefont {Udby}},\ }\href {\doibase 10.1103/PhysRevB.96.174106}
  {\bibfield  {journal} {\bibinfo  {journal} {Phys. Rev. B}\ }\textbf {\bibinfo
  {volume} {96}},\ \bibinfo {pages} {174106} (\bibinfo {year}
  {2017})}\BibitemShut {NoStop}%
\bibitem [{\citenamefont {Poccia}\ \emph {et~al.}(2012)\citenamefont {Poccia},
  \citenamefont {Ricci}, \citenamefont {Campi}, \citenamefont {Fratini},
  \citenamefont {Puri}, \citenamefont {Gioacchino}, \citenamefont {Marcelli},
  \citenamefont {Reynolds}, \citenamefont {Burghammer}, \citenamefont {Saini},
  \citenamefont {Aeppli},\ and\ \citenamefont {Bianconi}}]{Poccia2012}%
  \BibitemOpen
  \bibfield  {author} {\bibinfo {author} {\bibfnamefont {N.}~\bibnamefont
  {Poccia}}, \bibinfo {author} {\bibfnamefont {A.}~\bibnamefont {Ricci}},
  \bibinfo {author} {\bibfnamefont {G.}~\bibnamefont {Campi}}, \bibinfo
  {author} {\bibfnamefont {M.}~\bibnamefont {Fratini}}, \bibinfo {author}
  {\bibfnamefont {A.}~\bibnamefont {Puri}}, \bibinfo {author} {\bibfnamefont
  {D.~D.}\ \bibnamefont {Gioacchino}}, \bibinfo {author} {\bibfnamefont
  {A.}~\bibnamefont {Marcelli}}, \bibinfo {author} {\bibfnamefont
  {M.}~\bibnamefont {Reynolds}}, \bibinfo {author} {\bibfnamefont
  {M.}~\bibnamefont {Burghammer}}, \bibinfo {author} {\bibfnamefont {N.~L.}\
  \bibnamefont {Saini}}, \bibinfo {author} {\bibfnamefont {G.}~\bibnamefont
  {Aeppli}}, \ and\ \bibinfo {author} {\bibfnamefont {A.}~\bibnamefont
  {Bianconi}},\ }\href {\doibase 10.1073/pnas.1208492109} {\bibfield  {journal}
  {\bibinfo  {journal} {PNAS}\ }\textbf {\bibinfo {volume} {109}},\ \bibinfo
  {pages} {15685} (\bibinfo {year} {2012})}\BibitemShut {NoStop}%
\bibitem [{\citenamefont {Kaneshita}\ \emph {et~al.}(2002)\citenamefont
  {Kaneshita}, \citenamefont {Ichioka},\ and\ \citenamefont
  {Machida}}]{Kaneshita2002}%
  \BibitemOpen
  \bibfield  {author} {\bibinfo {author} {\bibfnamefont {E.}~\bibnamefont
  {Kaneshita}}, \bibinfo {author} {\bibfnamefont {M.}~\bibnamefont {Ichioka}},
  \ and\ \bibinfo {author} {\bibfnamefont {K.}~\bibnamefont {Machida}},\ }\href
  {\doibase 10.1103/PhysRevLett.88.115501} {\bibfield  {journal} {\bibinfo
  {journal} {Phys. Rev. Lett.}\ }\textbf {\bibinfo {volume} {88}},\ \bibinfo
  {pages} {115501} (\bibinfo {year} {2002})}\BibitemShut {NoStop}%
\bibitem [{\citenamefont {Udby}\ \emph {et~al.}(2013)\citenamefont {Udby},
  \citenamefont {Larsen}, \citenamefont {Christensen}, \citenamefont {Boehm},
  \citenamefont {Niedermayer}, \citenamefont {Mohottala}, \citenamefont
  {Jensen}, \citenamefont {{Toft-Petersen}}, \citenamefont {Chou},
  \citenamefont {Andersen}, \citenamefont {Lefmann},\ and\ \citenamefont
  {Wells}}]{Udby2013}%
  \BibitemOpen
  \bibfield  {author} {\bibinfo {author} {\bibfnamefont {L.}~\bibnamefont
  {Udby}}, \bibinfo {author} {\bibfnamefont {J.}~\bibnamefont {Larsen}},
  \bibinfo {author} {\bibfnamefont {N.~B.}\ \bibnamefont {Christensen}},
  \bibinfo {author} {\bibfnamefont {M.}~\bibnamefont {Boehm}}, \bibinfo
  {author} {\bibfnamefont {C.}~\bibnamefont {Niedermayer}}, \bibinfo {author}
  {\bibfnamefont {H.~E.}\ \bibnamefont {Mohottala}}, \bibinfo {author}
  {\bibfnamefont {T.~B.~S.}\ \bibnamefont {Jensen}}, \bibinfo {author}
  {\bibfnamefont {R.}~\bibnamefont {{Toft-Petersen}}}, \bibinfo {author}
  {\bibfnamefont {F.~C.}\ \bibnamefont {Chou}}, \bibinfo {author}
  {\bibfnamefont {N.~H.}\ \bibnamefont {Andersen}}, \bibinfo {author}
  {\bibfnamefont {K.}~\bibnamefont {Lefmann}}, \ and\ \bibinfo {author}
  {\bibfnamefont {B.~O.}\ \bibnamefont {Wells}},\ }\href {\doibase
  10.1103/PhysRevLett.111.227001} {\bibfield  {journal} {\bibinfo  {journal}
  {Phys. Rev. Lett.}\ }\textbf {\bibinfo {volume} {111}},\ \bibinfo {pages}
  {227001} (\bibinfo {year} {2013})}\BibitemShut {NoStop}%
\bibitem [{\citenamefont {Hirota}(2001)}]{Hirota2001}%
  \BibitemOpen
  \bibfield  {author} {\bibinfo {author} {\bibfnamefont {K.}~\bibnamefont
  {Hirota}},\ }\href {\doibase 10.1016/S0921-4534(01)00195-2} {\bibfield
  {journal} {\bibinfo  {journal} {Physica C: Superconductivity}\ }\textbf
  {\bibinfo {volume} {357-360}},\ \bibinfo {pages} {61} (\bibinfo {year}
  {2001})}\BibitemShut {NoStop}%
\bibitem [{\citenamefont {Julien}(2003)}]{Julien2003}%
  \BibitemOpen
  \bibfield  {author} {\bibinfo {author} {\bibfnamefont {M.~H.}\ \bibnamefont
  {Julien}},\ }\href {\doibase 10.1016/S0921-4526(02)01997-X} {\bibfield
  {journal} {\bibinfo  {journal} {Physica B: Condensed Matter}\ }\bibinfo
  {series} {Proceedings of the 23rd {{International Conference}} on {{Low
  Temperature Physics}}},\ \textbf {\bibinfo {volume} {329-333}},\ \bibinfo
  {pages} {693} (\bibinfo {year} {2003})}\BibitemShut {NoStop}%
\bibitem [{\citenamefont {Lee}\ \emph {et~al.}(2004)\citenamefont {Lee},
  \citenamefont {Chou}, \citenamefont {Tewary}, \citenamefont {Kastner},
  \citenamefont {Lee},\ and\ \citenamefont {Birgeneau}}]{Lee2004}%
  \BibitemOpen
  \bibfield  {author} {\bibinfo {author} {\bibfnamefont {Y.~S.}\ \bibnamefont
  {Lee}}, \bibinfo {author} {\bibfnamefont {F.~C.}\ \bibnamefont {Chou}},
  \bibinfo {author} {\bibfnamefont {A.}~\bibnamefont {Tewary}}, \bibinfo
  {author} {\bibfnamefont {M.~A.}\ \bibnamefont {Kastner}}, \bibinfo {author}
  {\bibfnamefont {S.~H.}\ \bibnamefont {Lee}}, \ and\ \bibinfo {author}
  {\bibfnamefont {R.~J.}\ \bibnamefont {Birgeneau}},\ }\href {\doibase
  10.1103/PhysRevB.69.020502} {\bibfield  {journal} {\bibinfo  {journal} {Phys.
  Rev. B}\ }\textbf {\bibinfo {volume} {69}},\ \bibinfo {pages} {020502}
  (\bibinfo {year} {2004})}\BibitemShut {NoStop}%
\bibitem [{\citenamefont {Holm-Dahlin}\ \emph {et~al.}(2019)\citenamefont
  {Holm-Dahlin}, \citenamefont {Larsen}, \citenamefont {Jacobsen},
  \citenamefont {R\o{}mer}, \citenamefont {\c{T}u\c{t}ueanu}, \citenamefont
  {Ahmad}, \citenamefont {Grivel}, \citenamefont {Goko}, \citenamefont
  {Scheuermann}, \citenamefont {v~Zimmermann}, \citenamefont {Boehm},
  \citenamefont {Steffens}, \citenamefont {Conder}, \citenamefont
  {Niedermayer}, \citenamefont {Pedersen}, \citenamefont {Christensen},
  \citenamefont {Emery}, \citenamefont {Wells}, \citenamefont {Lefmann},\ and\
  \citenamefont {Udby}}]{Holm2019}%
  \BibitemOpen
  \bibfield  {author} {\bibinfo {author} {\bibfnamefont {S.}~\bibnamefont
  {Holm-Dahlin}}, \bibinfo {author} {\bibfnamefont {J.}~\bibnamefont {Larsen}},
  \bibinfo {author} {\bibfnamefont {H.}~\bibnamefont {Jacobsen}}, \bibinfo
  {author} {\bibfnamefont {A.~T.}\ \bibnamefont {R\o{}mer}}, \bibinfo {author}
  {\bibfnamefont {A.-E.}\ \bibnamefont {\c{T}u\c{t}ueanu}}, \bibinfo {author}
  {\bibfnamefont {M.}~\bibnamefont {Ahmad}}, \bibinfo {author} {\bibfnamefont
  {J.-C.}\ \bibnamefont {Grivel}}, \bibinfo {author} {\bibfnamefont
  {T.}~\bibnamefont {Goko}}, \bibinfo {author} {\bibfnamefont {R.}~\bibnamefont
  {Scheuermann}}, \bibinfo {author} {\bibfnamefont {M.}~\bibnamefont
  {v~Zimmermann}}, \bibinfo {author} {\bibfnamefont {M.}~\bibnamefont {Boehm}},
  \bibinfo {author} {\bibfnamefont {P.}~\bibnamefont {Steffens}}, \bibinfo
  {author} {\bibfnamefont {K.}~\bibnamefont {Conder}}, \bibinfo {author}
  {\bibfnamefont {C.}~\bibnamefont {Niedermayer}}, \bibinfo {author}
  {\bibfnamefont {K.~S.}\ \bibnamefont {Pedersen}}, \bibinfo {author}
  {\bibfnamefont {N.~B.}\ \bibnamefont {Christensen}}, \bibinfo {author}
  {\bibfnamefont {S.~B.}\ \bibnamefont {Emery}}, \bibinfo {author}
  {\bibfnamefont {B.~O.}\ \bibnamefont {Wells}}, \bibinfo {author}
  {\bibfnamefont {K.}~\bibnamefont {Lefmann}}, \ and\ \bibinfo {author}
  {\bibfnamefont {L.}~\bibnamefont {Udby}},\ }\href@noop {} {\bibfield
  {journal} {\bibinfo  {journal} {Manuscript in preparation}\ } (\bibinfo
  {year} {2019})}\BibitemShut {NoStop}%
\bibitem [{\citenamefont {Tejsner}\ \emph
  {et~al.}(2018{\natexlab{a}})\citenamefont {Tejsner}, \citenamefont {Piovano},
  \citenamefont {\c{T}u\c{t}ueanu}, \citenamefont {Boehm},\ and\ \citenamefont
  {Udby}}]{dispersiondata}%
  \BibitemOpen
  \bibfield  {author} {\bibinfo {author} {\bibfnamefont {T.}~\bibnamefont
  {Tejsner}}, \bibinfo {author} {\bibfnamefont {A.}~\bibnamefont {Piovano}},
  \bibinfo {author} {\bibfnamefont {A.-E.}\ \bibnamefont {\c{T}u\c{t}ueanu}},
  \bibinfo {author} {\bibfnamefont {M.}~\bibnamefont {Boehm}}, \ and\ \bibinfo
  {author} {\bibfnamefont {L.}~\bibnamefont {Udby}},\ }\href {\doibase
  doi:10.5291/ILL-DATA.7-01-458} {\bibfield  {journal} {\bibinfo  {journal}
  {Institut Laue-Langevin (ILL)}\ } (\bibinfo {year} {2018}{\natexlab{a}}),\
  doi:10.5291/ILL-DATA.7-01-458}\BibitemShut {NoStop}%
\bibitem [{\citenamefont {Tejsner}\ \emph
  {et~al.}(2018{\natexlab{b}})\citenamefont {Tejsner}, \citenamefont {Piovano},
  \citenamefont {\c{T}u\c{t}ueanu}, \citenamefont {Boehm},\ and\ \citenamefont
  {Udby}}]{fielddata}%
  \BibitemOpen
  \bibfield  {author} {\bibinfo {author} {\bibfnamefont {T.}~\bibnamefont
  {Tejsner}}, \bibinfo {author} {\bibfnamefont {A.}~\bibnamefont {Piovano}},
  \bibinfo {author} {\bibfnamefont {A.-E.}\ \bibnamefont {\c{T}u\c{t}ueanu}},
  \bibinfo {author} {\bibfnamefont {M.}~\bibnamefont {Boehm}}, \ and\ \bibinfo
  {author} {\bibfnamefont {L.}~\bibnamefont {Udby}},\ }\href {\doibase
  doi:10.5291/ILL-DATA.7-01-474} {\bibfield  {journal} {\bibinfo  {journal}
  {Institut Laue-Langevin (ILL)}\ } (\bibinfo {year} {2018}{\natexlab{b}}),\
  doi:10.5291/ILL-DATA.7-01-474}\BibitemShut {NoStop}%
\bibitem [{\citenamefont {F\aa{}k}\ and\ \citenamefont
  {Dorner}(1997)}]{Fak1997}%
  \BibitemOpen
  \bibfield  {author} {\bibinfo {author} {\bibfnamefont {B.}~\bibnamefont
  {F\aa{}k}}\ and\ \bibinfo {author} {\bibfnamefont {B.}~\bibnamefont
  {Dorner}},\ }\href {\doibase 10.1016/S0921-4526(97)00121-X} {\bibfield
  {journal} {\bibinfo  {journal} {Physica B: Condensed Matter}\ }\bibinfo
  {series} {Proceedings of the {{First European Conference}} on {{Neutron
  Scattering}}},\ \textbf {\bibinfo {volume} {234-236}},\ \bibinfo {pages}
  {1107} (\bibinfo {year} {1997})}\BibitemShut {NoStop}%
\bibitem [{\citenamefont {Giustino}\ \emph {et~al.}(2008)\citenamefont
  {Giustino}, \citenamefont {Cohen},\ and\ \citenamefont
  {Louie}}]{Giustino2008}%
  \BibitemOpen
  \bibfield  {author} {\bibinfo {author} {\bibfnamefont {F.}~\bibnamefont
  {Giustino}}, \bibinfo {author} {\bibfnamefont {M.~L.}\ \bibnamefont {Cohen}},
  \ and\ \bibinfo {author} {\bibfnamefont {S.~G.}\ \bibnamefont {Louie}},\
  }\href {\doibase 10.1038/nature06874} {\bibfield  {journal} {\bibinfo
  {journal} {Nature}\ }\textbf {\bibinfo {volume} {452}},\ \bibinfo {pages}
  {975} (\bibinfo {year} {2008})}\BibitemShut {NoStop}%
\bibitem [{\citenamefont {Reznik}(2012)}]{Reznik2012}%
  \BibitemOpen
  \bibfield  {author} {\bibinfo {author} {\bibfnamefont {D.}~\bibnamefont
  {Reznik}},\ }\href {\doibase 10.1016/j.physc.2012.01.024} {\bibfield
  {journal} {\bibinfo  {journal} {Physica C: Superconductivity}\ }\bibinfo
  {series} {Stripes and {{Electronic Liquid Crystals}} in {{Strongly Correlated
  Materials}}},\ \textbf {\bibinfo {volume} {481}},\ \bibinfo {pages} {75}
  (\bibinfo {year} {2012})}\BibitemShut {NoStop}%
\bibitem [{\citenamefont {Reznik}\ \emph {et~al.}(2008)\citenamefont {Reznik},
  \citenamefont {Fukuda}, \citenamefont {Lamago}, \citenamefont {Baron},
  \citenamefont {Tsutsui}, \citenamefont {Fujita},\ and\ \citenamefont
  {Yamada}}]{Reznik2008b}%
  \BibitemOpen
  \bibfield  {author} {\bibinfo {author} {\bibfnamefont {D.}~\bibnamefont
  {Reznik}}, \bibinfo {author} {\bibfnamefont {T.}~\bibnamefont {Fukuda}},
  \bibinfo {author} {\bibfnamefont {D.}~\bibnamefont {Lamago}}, \bibinfo
  {author} {\bibfnamefont {A.~Q.~R.}\ \bibnamefont {Baron}}, \bibinfo {author}
  {\bibfnamefont {S.}~\bibnamefont {Tsutsui}}, \bibinfo {author} {\bibfnamefont
  {M.}~\bibnamefont {Fujita}}, \ and\ \bibinfo {author} {\bibfnamefont
  {K.}~\bibnamefont {Yamada}},\ }\href {\doibase 10.1016/j.jpcs.2008.06.027}
  {\bibfield  {journal} {\bibinfo  {journal} {Journal of Physics and Chemistry
  of Solids}\ }\bibinfo {series} {{{SNS2007}}},\ \textbf {\bibinfo {volume}
  {69}},\ \bibinfo {pages} {3103} (\bibinfo {year} {2008})}\BibitemShut
  {NoStop}%
\bibitem [{\citenamefont {Park}\ \emph {et~al.}(2013)\citenamefont {Park},
  \citenamefont {Cao}, \citenamefont {Wang}, \citenamefont {Fujita},
  \citenamefont {Yamada}, \citenamefont {Mo}, \citenamefont {Dessau},\ and\
  \citenamefont {Reznik}}]{Park2013}%
  \BibitemOpen
  \bibfield  {author} {\bibinfo {author} {\bibfnamefont {S.~R.}\ \bibnamefont
  {Park}}, \bibinfo {author} {\bibfnamefont {Y.}~\bibnamefont {Cao}}, \bibinfo
  {author} {\bibfnamefont {Q.}~\bibnamefont {Wang}}, \bibinfo {author}
  {\bibfnamefont {M.}~\bibnamefont {Fujita}}, \bibinfo {author} {\bibfnamefont
  {K.}~\bibnamefont {Yamada}}, \bibinfo {author} {\bibfnamefont {S.-K.}\
  \bibnamefont {Mo}}, \bibinfo {author} {\bibfnamefont {D.~S.}\ \bibnamefont
  {Dessau}}, \ and\ \bibinfo {author} {\bibfnamefont {D.}~\bibnamefont
  {Reznik}},\ }\href {\doibase 10.1103/PhysRevB.88.220503} {\bibfield
  {journal} {\bibinfo  {journal} {Phys. Rev. B}\ }\textbf {\bibinfo {volume}
  {88}},\ \bibinfo {pages} {220503} (\bibinfo {year} {2013})}\BibitemShut
  {NoStop}%
\bibitem [{\citenamefont {Garcia}\ and\ \citenamefont
  {Lanzara}(2010)}]{Garcia2010}%
  \BibitemOpen
  \bibfield  {author} {\bibinfo {author} {\bibfnamefont {D.~R.}\ \bibnamefont
  {Garcia}}\ and\ \bibinfo {author} {\bibfnamefont {A.}~\bibnamefont
  {Lanzara}},\ }\href {\doibase 10.1155/2010/807412} {\bibfield  {journal}
  {\bibinfo  {journal} {Advances in Condensed Matter Physics}\ }\textbf
  {\bibinfo {volume} {2010}} (\bibinfo {year} {2010}),\
  10.1155/2010/807412}\BibitemShut {NoStop}%
\bibitem [{\citenamefont {Zhang}\ \emph {et~al.}(2018)\citenamefont {Zhang},
  \citenamefont {Sutarto}, \citenamefont {He}, \citenamefont {Chou},
  \citenamefont {Udby}, \citenamefont {Holm}, \citenamefont {Zhu},
  \citenamefont {Hines}, \citenamefont {Budnick},\ and\ \citenamefont
  {Wells}}]{Zhang2018}%
  \BibitemOpen
  \bibfield  {author} {\bibinfo {author} {\bibfnamefont {Z.}~\bibnamefont
  {Zhang}}, \bibinfo {author} {\bibfnamefont {R.}~\bibnamefont {Sutarto}},
  \bibinfo {author} {\bibfnamefont {F.}~\bibnamefont {He}}, \bibinfo {author}
  {\bibfnamefont {F.~C.}\ \bibnamefont {Chou}}, \bibinfo {author}
  {\bibfnamefont {L.}~\bibnamefont {Udby}}, \bibinfo {author} {\bibfnamefont
  {S.~L.}\ \bibnamefont {Holm}}, \bibinfo {author} {\bibfnamefont {Z.~H.}\
  \bibnamefont {Zhu}}, \bibinfo {author} {\bibfnamefont {W.~A.}\ \bibnamefont
  {Hines}}, \bibinfo {author} {\bibfnamefont {J.~I.}\ \bibnamefont {Budnick}},
  \ and\ \bibinfo {author} {\bibfnamefont {B.~O.}\ \bibnamefont {Wells}},\
  }\href {\doibase 10.1103/PhysRevLett.121.067602} {\bibfield  {journal}
  {\bibinfo  {journal} {Phys. Rev. Lett.}\ }\textbf {\bibinfo {volume} {121}},\
  \bibinfo {pages} {067602} (\bibinfo {year} {2018})}\BibitemShut {NoStop}%
\bibitem [{Udb(2019)}]{Udby2019}%
  \BibitemOpen
  \href@noop {} {\bibfield  {journal} {\bibinfo  {journal} {Manuscript in
  preparation}\ } (\bibinfo {year} {2019})}\BibitemShut {NoStop}%
\bibitem [{\citenamefont {Reznik}\ \emph {et~al.}(2016)\citenamefont {Reznik},
  \citenamefont {Parshall}, \citenamefont {Park}, \citenamefont {Lynn},\ and\
  \citenamefont {Wolf}}]{Reznik2016}%
  \BibitemOpen
  \bibfield  {author} {\bibinfo {author} {\bibfnamefont {D.}~\bibnamefont
  {Reznik}}, \bibinfo {author} {\bibfnamefont {D.}~\bibnamefont {Parshall}},
  \bibinfo {author} {\bibfnamefont {S.~R.}\ \bibnamefont {Park}}, \bibinfo
  {author} {\bibfnamefont {J.~W.}\ \bibnamefont {Lynn}}, \ and\ \bibinfo
  {author} {\bibfnamefont {T.}~\bibnamefont {Wolf}},\ }\href {\doibase
  10.1007/s10948-015-3293-1} {\bibfield  {journal} {\bibinfo  {journal} {J
  Supercond Nov Magn}\ }\textbf {\bibinfo {volume} {29}},\ \bibinfo {pages}
  {643} (\bibinfo {year} {2016})}\BibitemShut {NoStop}%
\bibitem [{\citenamefont {Fujita}\ \emph {et~al.}(2004)\citenamefont {Fujita},
  \citenamefont {Goka}, \citenamefont {Yamada}, \citenamefont {Tranquada},\
  and\ \citenamefont {Regnault}}]{Fujita2004}%
  \BibitemOpen
  \bibfield  {author} {\bibinfo {author} {\bibfnamefont {M.}~\bibnamefont
  {Fujita}}, \bibinfo {author} {\bibfnamefont {H.}~\bibnamefont {Goka}},
  \bibinfo {author} {\bibfnamefont {K.}~\bibnamefont {Yamada}}, \bibinfo
  {author} {\bibfnamefont {J.~M.}\ \bibnamefont {Tranquada}}, \ and\ \bibinfo
  {author} {\bibfnamefont {L.~P.}\ \bibnamefont {Regnault}},\ }\href {\doibase
  10.1103/PhysRevB.70.104517} {\bibfield  {journal} {\bibinfo  {journal} {Phys.
  Rev. B}\ }\textbf {\bibinfo {volume} {70}},\ \bibinfo {pages} {104517}
  (\bibinfo {year} {2004})}\BibitemShut {NoStop}%
\bibitem [{\citenamefont {Jacobsen}\ \emph {et~al.}(2018)\citenamefont
  {Jacobsen}, \citenamefont {Holm}, \citenamefont {L{\u a}c{\u a}tu{\c s}u},
  \citenamefont {R\o{}mer}, \citenamefont {Bertelsen}, \citenamefont {Boehm},
  \citenamefont {{Toft-Petersen}}, \citenamefont {Grivel}, \citenamefont
  {Emery}, \citenamefont {Udby}, \citenamefont {Wells},\ and\ \citenamefont
  {Lefmann}}]{Jacobsen2018}%
  \BibitemOpen
  \bibfield  {author} {\bibinfo {author} {\bibfnamefont {H.}~\bibnamefont
  {Jacobsen}}, \bibinfo {author} {\bibfnamefont {S.~L.}\ \bibnamefont {Holm}},
  \bibinfo {author} {\bibfnamefont {M.-E.}\ \bibnamefont {L{\u a}c{\u a}tu{\c
  s}u}}, \bibinfo {author} {\bibfnamefont {A.~T.}\ \bibnamefont {R\o{}mer}},
  \bibinfo {author} {\bibfnamefont {M.}~\bibnamefont {Bertelsen}}, \bibinfo
  {author} {\bibfnamefont {M.}~\bibnamefont {Boehm}}, \bibinfo {author}
  {\bibfnamefont {R.}~\bibnamefont {{Toft-Petersen}}}, \bibinfo {author}
  {\bibfnamefont {J.-C.}\ \bibnamefont {Grivel}}, \bibinfo {author}
  {\bibfnamefont {S.~B.}\ \bibnamefont {Emery}}, \bibinfo {author}
  {\bibfnamefont {L.}~\bibnamefont {Udby}}, \bibinfo {author} {\bibfnamefont
  {B.~O.}\ \bibnamefont {Wells}}, \ and\ \bibinfo {author} {\bibfnamefont
  {K.}~\bibnamefont {Lefmann}},\ }\href {\doibase
  10.1103/PhysRevLett.120.037003} {\bibfield  {journal} {\bibinfo  {journal}
  {Phys. Rev. Lett.}\ }\textbf {\bibinfo {volume} {120}},\ \bibinfo {pages}
  {037003} (\bibinfo {year} {2018})}\BibitemShut {NoStop}%
\bibitem [{\citenamefont {Fradkin}\ \emph {et~al.}(2015)\citenamefont
  {Fradkin}, \citenamefont {Kivelson},\ and\ \citenamefont
  {Tranquada}}]{Fradkin2015}%
  \BibitemOpen
  \bibfield  {author} {\bibinfo {author} {\bibfnamefont {E.}~\bibnamefont
  {Fradkin}}, \bibinfo {author} {\bibfnamefont {S.~A.}\ \bibnamefont
  {Kivelson}}, \ and\ \bibinfo {author} {\bibfnamefont {J.~M.}\ \bibnamefont
  {Tranquada}},\ }\href {\doibase 10.1103/RevModPhys.87.457} {\bibfield
  {journal} {\bibinfo  {journal} {Rev. Mod. Phys.}\ }\textbf {\bibinfo {volume}
  {87}},\ \bibinfo {pages} {457} (\bibinfo {year} {2015})}\BibitemShut
  {NoStop}%
\bibitem [{\citenamefont {Kivelson}\ \emph {et~al.}(1998)\citenamefont
  {Kivelson}, \citenamefont {Fradkin},\ and\ \citenamefont
  {Emery}}]{Kivelson1998}%
  \BibitemOpen
  \bibfield  {author} {\bibinfo {author} {\bibfnamefont {S.~A.}\ \bibnamefont
  {Kivelson}}, \bibinfo {author} {\bibfnamefont {E.}~\bibnamefont {Fradkin}}, \
  and\ \bibinfo {author} {\bibfnamefont {V.~J.}\ \bibnamefont {Emery}},\ }\href
  {\doibase 10.1038/31177} {\bibfield  {journal} {\bibinfo  {journal} {Nature}\
  }\textbf {\bibinfo {volume} {393}},\ \bibinfo {pages} {550} (\bibinfo {year}
  {1998})}\BibitemShut {NoStop}%
\end{thebibliography}%


\begin{thebibliography}{13}%
\makeatletter
\providecommand \@ifxundefined [1]{%
 \@ifx{#1\undefined}
}%
\providecommand \@ifnum [1]{%
 \ifnum #1\expandafter \@firstoftwo
 \else \expandafter \@secondoftwo
 \fi
}%
\providecommand \@ifx [1]{%
 \ifx #1\expandafter \@firstoftwo
 \else \expandafter \@secondoftwo
 \fi
}%
\providecommand \natexlab [1]{#1}%
\providecommand \enquote  [1]{``#1''}%
\providecommand \bibnamefont  [1]{#1}%
\providecommand \bibfnamefont [1]{#1}%
\providecommand \citenamefont [1]{#1}%
\providecommand \href@noop [0]{\@secondoftwo}%
\providecommand \href [0]{\begingroup \@sanitize@url \@href}%
\providecommand \@href[1]{\@@startlink{#1}\@@href}%
\providecommand \@@href[1]{\endgroup#1\@@endlink}%
\providecommand \@sanitize@url [0]{\catcode `\\12\catcode `\$12\catcode
  `\&12\catcode `\#12\catcode `\^12\catcode `\_12\catcode `\%12\relax}%
\providecommand \@@startlink[1]{}%
\providecommand \@@endlink[0]{}%
\providecommand \url  [0]{\begingroup\@sanitize@url \@url }%
\providecommand \@url [1]{\endgroup\@href {#1}{\urlprefix }}%
\providecommand \urlprefix  [0]{URL }%
\providecommand \Eprint [0]{\href }%
\providecommand \doibase [0]{http://dx.doi.org/}%
\providecommand \selectlanguage [0]{\@gobble}%
\providecommand \bibinfo  [0]{\@secondoftwo}%
\providecommand \bibfield  [0]{\@secondoftwo}%
\providecommand \translation [1]{[#1]}%
\providecommand \BibitemOpen [0]{}%
\providecommand \bibitemStop [0]{}%
\providecommand \bibitemNoStop [0]{.\EOS\space}%
\providecommand \EOS [0]{\spacefactor3000\relax}%
\providecommand \BibitemShut  [1]{\csname bibitem#1\endcsname}%
\let\auto@bib@innerbib\@empty
\bibitem [{\citenamefont {Jacobsen}\ \emph {et~al.}(2018)\citenamefont
  {Jacobsen}, \citenamefont {Holm}, \citenamefont {L{\u a}c{\u a}tu{\c s}u},
  \citenamefont {R{\o}mer}, \citenamefont {Bertelsen}, \citenamefont {Boehm},
  \citenamefont {{Toft-Petersen}}, \citenamefont {Grivel}, \citenamefont
  {Emery}, \citenamefont {Udby}, \citenamefont {Wells},\ and\ \citenamefont
  {Lefmann}}]{Jacobsen2018}%
  \BibitemOpen
  \bibfield  {author} {\bibinfo {author} {\bibfnamefont {H.}~\bibnamefont
  {Jacobsen}}, \bibinfo {author} {\bibfnamefont {S.~L.}\ \bibnamefont {Holm}},
  \bibinfo {author} {\bibfnamefont {M.-E.}\ \bibnamefont {L{\u a}c{\u a}tu{\c
  s}u}}, \bibinfo {author} {\bibfnamefont {A.~T.}\ \bibnamefont {R{\o}mer}},
  \bibinfo {author} {\bibfnamefont {M.}~\bibnamefont {Bertelsen}}, \bibinfo
  {author} {\bibfnamefont {M.}~\bibnamefont {Boehm}}, \bibinfo {author}
  {\bibfnamefont {R.}~\bibnamefont {{Toft-Petersen}}}, \bibinfo {author}
  {\bibfnamefont {J.-C.}\ \bibnamefont {Grivel}}, \bibinfo {author}
  {\bibfnamefont {S.~B.}\ \bibnamefont {Emery}}, \bibinfo {author}
  {\bibfnamefont {L.}~\bibnamefont {Udby}}, \bibinfo {author} {\bibfnamefont
  {B.~O.}\ \bibnamefont {Wells}}, \ and\ \bibinfo {author} {\bibfnamefont
  {K.}~\bibnamefont {Lefmann}},\ }\href {\doibase
  10.1103/PhysRevLett.120.037003} {\bibfield  {journal} {\bibinfo  {journal}
  {Phys. Rev. Lett.}\ }\textbf {\bibinfo {volume} {120}},\ \bibinfo {pages}
  {037003} (\bibinfo {year} {2018})}\BibitemShut {NoStop}%
\bibitem [{\citenamefont {McQueeney}\ \emph {et~al.}(1999)\citenamefont
  {McQueeney}, \citenamefont {Petrov}, \citenamefont {Egami}, \citenamefont
  {Yethiraj}, \citenamefont {Shirane},\ and\ \citenamefont
  {Endoh}}]{McQueeney1999}%
  \BibitemOpen
  \bibfield  {author} {\bibinfo {author} {\bibfnamefont {R.~J.}\ \bibnamefont
  {McQueeney}}, \bibinfo {author} {\bibfnamefont {Y.}~\bibnamefont {Petrov}},
  \bibinfo {author} {\bibfnamefont {T.}~\bibnamefont {Egami}}, \bibinfo
  {author} {\bibfnamefont {M.}~\bibnamefont {Yethiraj}}, \bibinfo {author}
  {\bibfnamefont {G.}~\bibnamefont {Shirane}}, \ and\ \bibinfo {author}
  {\bibfnamefont {Y.}~\bibnamefont {Endoh}},\ }\href {\doibase
  10.1103/PhysRevLett.82.628} {\bibfield  {journal} {\bibinfo  {journal} {Phys.
  Rev. Lett.}\ }\textbf {\bibinfo {volume} {82}},\ \bibinfo {pages} {628}
  (\bibinfo {year} {1999})}\BibitemShut {NoStop}%
\bibitem [{\citenamefont {Reznik}\ \emph {et~al.}(2007)\citenamefont {Reznik},
  \citenamefont {Pintschovius}, \citenamefont {Fujita}, \citenamefont {Yamada},
  \citenamefont {Gu},\ and\ \citenamefont {Tranquada}}]{Reznik2007}%
  \BibitemOpen
  \bibfield  {author} {\bibinfo {author} {\bibfnamefont {D.}~\bibnamefont
  {Reznik}}, \bibinfo {author} {\bibfnamefont {L.}~\bibnamefont
  {Pintschovius}}, \bibinfo {author} {\bibfnamefont {M.}~\bibnamefont
  {Fujita}}, \bibinfo {author} {\bibfnamefont {K.}~\bibnamefont {Yamada}},
  \bibinfo {author} {\bibfnamefont {G.~D.}\ \bibnamefont {Gu}}, \ and\ \bibinfo
  {author} {\bibfnamefont {J.~M.}\ \bibnamefont {Tranquada}},\ }\href {\doibase
  10.1007/s10909-007-9318-9} {\bibfield  {journal} {\bibinfo  {journal} {J Low
  Temp Phys}\ }\textbf {\bibinfo {volume} {147}},\ \bibinfo {pages} {353}
  (\bibinfo {year} {2007})}\BibitemShut {NoStop}%
\bibitem [{\citenamefont {Shirane}\ \emph {et~al.}(2002)\citenamefont
  {Shirane}, \citenamefont {Shapiro},\ and\ \citenamefont
  {Tranquada}}]{Shirane2002}%
  \BibitemOpen
  \bibfield  {author} {\bibinfo {author} {\bibfnamefont {G.}~\bibnamefont
  {Shirane}}, \bibinfo {author} {\bibfnamefont {S.~M.}\ \bibnamefont
  {Shapiro}}, \ and\ \bibinfo {author} {\bibfnamefont {J.~M.}\ \bibnamefont
  {Tranquada}},\ }\href@noop {} {\emph {\bibinfo {title} {Neutron
  {{Scattering}} with a {{Triple}}-{{Axis Spectrometer}}: {{Basic
  Techniques}}}}}\ (\bibinfo  {publisher} {{Cambridge University Press}},\
  \bibinfo {year} {2002})\BibitemShut {NoStop}%
\bibitem [{\citenamefont {Kresse}\ and\ \citenamefont
  {Hafner}(1993)}]{Kresse1993}%
  \BibitemOpen
  \bibfield  {author} {\bibinfo {author} {\bibfnamefont {G.}~\bibnamefont
  {Kresse}}\ and\ \bibinfo {author} {\bibfnamefont {J.}~\bibnamefont
  {Hafner}},\ }\href {\doibase 10.1103/PhysRevB.48.13115} {\bibfield  {journal}
  {\bibinfo  {journal} {Phys. Rev. B}\ }\textbf {\bibinfo {volume} {48}},\
  \bibinfo {pages} {13115} (\bibinfo {year} {1993})}\BibitemShut {NoStop}%
\bibitem [{\citenamefont {Kresse}\ and\ \citenamefont
  {Furthm{\"u}ller}(1996{\natexlab{a}})}]{Kresse1996}%
  \BibitemOpen
  \bibfield  {author} {\bibinfo {author} {\bibfnamefont {G.}~\bibnamefont
  {Kresse}}\ and\ \bibinfo {author} {\bibfnamefont {J.}~\bibnamefont
  {Furthm{\"u}ller}},\ }\href {\doibase 10.1016/0927-0256(96)00008-0}
  {\bibfield  {journal} {\bibinfo  {journal} {Computational Materials Science}\
  }\textbf {\bibinfo {volume} {6}},\ \bibinfo {pages} {15} (\bibinfo {year}
  {1996}{\natexlab{a}})}\BibitemShut {NoStop}%
\bibitem [{\citenamefont {Kresse}\ and\ \citenamefont
  {Furthm{\"u}ller}(1996{\natexlab{b}})}]{Kresse1996a}%
  \BibitemOpen
  \bibfield  {author} {\bibinfo {author} {\bibfnamefont {G.}~\bibnamefont
  {Kresse}}\ and\ \bibinfo {author} {\bibfnamefont {J.}~\bibnamefont
  {Furthm{\"u}ller}},\ }\href {\doibase 10.1103/PhysRevB.54.11169} {\bibfield
  {journal} {\bibinfo  {journal} {Phys. Rev. B}\ }\textbf {\bibinfo {volume}
  {54}},\ \bibinfo {pages} {11169} (\bibinfo {year}
  {1996}{\natexlab{b}})}\BibitemShut {NoStop}%
\bibitem [{\citenamefont {Kresse}\ and\ \citenamefont
  {Joubert}(1999)}]{Kresse1999}%
  \BibitemOpen
  \bibfield  {author} {\bibinfo {author} {\bibfnamefont {G.}~\bibnamefont
  {Kresse}}\ and\ \bibinfo {author} {\bibfnamefont {D.}~\bibnamefont
  {Joubert}},\ }\href {\doibase 10.1103/PhysRevB.59.1758} {\bibfield  {journal}
  {\bibinfo  {journal} {Phys. Rev. B}\ }\textbf {\bibinfo {volume} {59}},\
  \bibinfo {pages} {1758} (\bibinfo {year} {1999})}\BibitemShut {NoStop}%
\bibitem [{\citenamefont {Csonka}\ \emph {et~al.}(2009)\citenamefont {Csonka},
  \citenamefont {Perdew}, \citenamefont {Ruzsinszky}, \citenamefont
  {Philipsen}, \citenamefont {Leb{\`e}gue}, \citenamefont {Paier},
  \citenamefont {Vydrov},\ and\ \citenamefont {{\'A}ngy{\'a}n}}]{Csonka2009}%
  \BibitemOpen
  \bibfield  {author} {\bibinfo {author} {\bibfnamefont {G.~I.}\ \bibnamefont
  {Csonka}}, \bibinfo {author} {\bibfnamefont {J.~P.}\ \bibnamefont {Perdew}},
  \bibinfo {author} {\bibfnamefont {A.}~\bibnamefont {Ruzsinszky}}, \bibinfo
  {author} {\bibfnamefont {P.~H.~T.}\ \bibnamefont {Philipsen}}, \bibinfo
  {author} {\bibfnamefont {S.}~\bibnamefont {Leb{\`e}gue}}, \bibinfo {author}
  {\bibfnamefont {J.}~\bibnamefont {Paier}}, \bibinfo {author} {\bibfnamefont
  {O.~A.}\ \bibnamefont {Vydrov}}, \ and\ \bibinfo {author} {\bibfnamefont
  {J.~G.}\ \bibnamefont {{\'A}ngy{\'a}n}},\ }\href {\doibase
  10.1103/PhysRevB.79.155107} {\bibfield  {journal} {\bibinfo  {journal} {Phys.
  Rev. B}\ }\textbf {\bibinfo {volume} {79}},\ \bibinfo {pages} {155107}
  (\bibinfo {year} {2009})}\BibitemShut {NoStop}%
\bibitem [{\citenamefont {Togo}\ and\ \citenamefont {Tanaka}(2015)}]{Togo2015}%
  \BibitemOpen
  \bibfield  {author} {\bibinfo {author} {\bibfnamefont {A.}~\bibnamefont
  {Togo}}\ and\ \bibinfo {author} {\bibfnamefont {I.}~\bibnamefont {Tanaka}},\
  }\href {\doibase 10.1016/j.scriptamat.2015.07.021} {\bibfield  {journal}
  {\bibinfo  {journal} {Scripta Materialia}\ }\textbf {\bibinfo {volume}
  {108}},\ \bibinfo {pages} {1} (\bibinfo {year} {2015})}\BibitemShut {NoStop}%
\bibitem [{\citenamefont {Matt}\ \emph {et~al.}(2018)\citenamefont {Matt},
  \citenamefont {Sutter}, \citenamefont {Cook}, \citenamefont {Sassa},
  \citenamefont {M{\aa}nsson}, \citenamefont {Tjernberg}, \citenamefont {Das},
  \citenamefont {Horio}, \citenamefont {Destraz}, \citenamefont {Fatuzzo},
  \citenamefont {Hauser}, \citenamefont {Shi}, \citenamefont {Kobayashi},
  \citenamefont {Strocov}, \citenamefont {Schmitt}, \citenamefont {Dudin},
  \citenamefont {Hoesch}, \citenamefont {Pyon}, \citenamefont {Takayama},
  \citenamefont {Takagi}, \citenamefont {Lipscombe}, \citenamefont {Hayden},
  \citenamefont {Kurosawa}, \citenamefont {Momono}, \citenamefont {Oda},
  \citenamefont {Neupert},\ and\ \citenamefont {Chang}}]{Matt2018}%
  \BibitemOpen
  \bibfield  {author} {\bibinfo {author} {\bibfnamefont {C.~E.}\ \bibnamefont
  {Matt}}, \bibinfo {author} {\bibfnamefont {D.}~\bibnamefont {Sutter}},
  \bibinfo {author} {\bibfnamefont {A.~M.}\ \bibnamefont {Cook}}, \bibinfo
  {author} {\bibfnamefont {Y.}~\bibnamefont {Sassa}}, \bibinfo {author}
  {\bibfnamefont {M.}~\bibnamefont {M{\aa}nsson}}, \bibinfo {author}
  {\bibfnamefont {O.}~\bibnamefont {Tjernberg}}, \bibinfo {author}
  {\bibfnamefont {L.}~\bibnamefont {Das}}, \bibinfo {author} {\bibfnamefont
  {M.}~\bibnamefont {Horio}}, \bibinfo {author} {\bibfnamefont
  {D.}~\bibnamefont {Destraz}}, \bibinfo {author} {\bibfnamefont {C.~G.}\
  \bibnamefont {Fatuzzo}}, \bibinfo {author} {\bibfnamefont {K.}~\bibnamefont
  {Hauser}}, \bibinfo {author} {\bibfnamefont {M.}~\bibnamefont {Shi}},
  \bibinfo {author} {\bibfnamefont {M.}~\bibnamefont {Kobayashi}}, \bibinfo
  {author} {\bibfnamefont {V.~N.}\ \bibnamefont {Strocov}}, \bibinfo {author}
  {\bibfnamefont {T.}~\bibnamefont {Schmitt}}, \bibinfo {author} {\bibfnamefont
  {P.}~\bibnamefont {Dudin}}, \bibinfo {author} {\bibfnamefont
  {M.}~\bibnamefont {Hoesch}}, \bibinfo {author} {\bibfnamefont
  {S.}~\bibnamefont {Pyon}}, \bibinfo {author} {\bibfnamefont {T.}~\bibnamefont
  {Takayama}}, \bibinfo {author} {\bibfnamefont {H.}~\bibnamefont {Takagi}},
  \bibinfo {author} {\bibfnamefont {O.~J.}\ \bibnamefont {Lipscombe}}, \bibinfo
  {author} {\bibfnamefont {S.~M.}\ \bibnamefont {Hayden}}, \bibinfo {author}
  {\bibfnamefont {T.}~\bibnamefont {Kurosawa}}, \bibinfo {author}
  {\bibfnamefont {N.}~\bibnamefont {Momono}}, \bibinfo {author} {\bibfnamefont
  {M.}~\bibnamefont {Oda}}, \bibinfo {author} {\bibfnamefont {T.}~\bibnamefont
  {Neupert}}, \ and\ \bibinfo {author} {\bibfnamefont {J.}~\bibnamefont
  {Chang}},\ }\href {\doibase 10.1038/s41467-018-03266-0} {\bibfield  {journal}
  {\bibinfo  {journal} {Nature Communications}\ }\textbf {\bibinfo {volume}
  {9}},\ \bibinfo {pages} {972} (\bibinfo {year} {2018})}\BibitemShut {NoStop}%
\bibitem [{\citenamefont {{\v S}aroun}\ and\ \citenamefont
  {Kulda}(1997)}]{Saroun1997}%
  \BibitemOpen
  \bibfield  {author} {\bibinfo {author} {\bibfnamefont {J.}~\bibnamefont {{\v
  S}aroun}}\ and\ \bibinfo {author} {\bibfnamefont {J.}~\bibnamefont {Kulda}},\
  }\href {\doibase 10.1016/S0921-4526(97)00037-9} {\bibfield  {journal}
  {\bibinfo  {journal} {Physica B: Condensed Matter}\ }\bibinfo {series}
  {Proceedings of the {{First European Conference}} on {{Neutron
  Scattering}}},\ \textbf {\bibinfo {volume} {234-236}},\ \bibinfo {pages}
  {1102} (\bibinfo {year} {1997})}\BibitemShut {NoStop}%
\bibitem [{\citenamefont {Mohottala}\ \emph {et~al.}(2006)\citenamefont
  {Mohottala}, \citenamefont {Wells}, \citenamefont {Budnick}, \citenamefont
  {Hines}, \citenamefont {Niedermayer}, \citenamefont {Udby}, \citenamefont
  {Bernhard}, \citenamefont {Moodenbaugh},\ and\ \citenamefont
  {Chou}}]{Mohottala2006}%
  \BibitemOpen
  \bibfield  {author} {\bibinfo {author} {\bibfnamefont {H.~E.}\ \bibnamefont
  {Mohottala}}, \bibinfo {author} {\bibfnamefont {B.~O.}\ \bibnamefont
  {Wells}}, \bibinfo {author} {\bibfnamefont {J.~I.}\ \bibnamefont {Budnick}},
  \bibinfo {author} {\bibfnamefont {W.~A.}\ \bibnamefont {Hines}}, \bibinfo
  {author} {\bibfnamefont {C.}~\bibnamefont {Niedermayer}}, \bibinfo {author}
  {\bibfnamefont {L.}~\bibnamefont {Udby}}, \bibinfo {author} {\bibfnamefont
  {C.}~\bibnamefont {Bernhard}}, \bibinfo {author} {\bibfnamefont {A.~R.}\
  \bibnamefont {Moodenbaugh}}, \ and\ \bibinfo {author} {\bibfnamefont {F.-C.}\
  \bibnamefont {Chou}},\ }\href {\doibase 10.1038/nmat1633} {\bibfield
  {journal} {\bibinfo  {journal} {Nature Materials}\ }\textbf {\bibinfo
  {volume} {5}},\ \bibinfo {pages} {377} (\bibinfo {year} {2006})}\BibitemShut
  {NoStop}%
\end{thebibliography}%

\end{document}